\documentclass[printer]{aa}
\usepackage[utf8]{inputenc}
\usepackage{amsmath}
\usepackage{amsfonts}
\usepackage{amssymb}
\usepackage{graphicx}
\usepackage{natbib}
\usepackage{nccmath}
\usepackage{hyperref}
\usepackage{xcolor}
\usepackage{balance}

\graphicspath{{lastfigures/}}

\bibpunct{(}{)}{;}{a}{}{,}

\author{A. Adrover-González\inst{1,2} \& J. Terradas\inst{1,2}}
\institute{Departamet de Física, Universitat de les Illes Balears, E-07122 Palma de Mallorca, Spain \and Institut d'Aplicacions Computacionals de Codi Comunitari (IAC\textsuperscript{3}), Universitat de les Illes Balears, E-07122 Palma de Mallorca, Spain}
\begin{document}
        
\title{3D numerical simulations of oscillations in solar prominences}
\date{} 
\abstract {Oscillations in solar prominences are a frequent phenomenon, and they have been the subject of many studies. A full understanding of the mechanisms that drive them and their attenuation has not been reached yet, however.} {We numerically investigate the periodicity and damping of transverse and longitudinal oscillations in a 3D model of a curtain-shaped prominence.} {We carried out a set of numerical simulations of vertical, transverse and longitudinal oscillations with the high-order finite-difference Pencil Code\textit{}. We solved the ideal magnetohydrodynamic (MHD) equations for a wide range of parameters, including the width ($w_x$) and density ($\rho_{p0}$) of the prominence, and the magnetic field strength ($B$) of the solar corona. We studied the periodicity and attenuation of the induced oscillations.} {We found that longitudinal oscillations can be fit with the pendulum model, whose restoring force is the field-aligned component of gravity, but other mechanisms such as pressure gradients may contribute to the movement. On the other hand, transverse oscillations are subject to magnetic forces. The analysis of the parametric survey shows, in agreement with observational studies, that the oscillation period ($P$) increases with the prominence width. For transverse oscillations we obtained that $P$ increases with density and decreases with $B$. For longitudinal oscillations we also found that $P$ increases with $\rho_{p0}$, but there are no variations with $B$. The attenuation of transverse oscillations was investigated by analysing the velocity distribution and computing the Alfvén continuum modes. We conclude that resonant absorption is the mean cause. Damping of longitudinal oscillations is due to some kind of shear numerical viscosity.} {Our model is a good approximation of a prominence body that nearly reproduces the observed oscillations. However, more realistic simulations that include other terms such as non-adiabatic processes or partially ionised plasmas are necessary to obtain better results.} 
\titlerunning{3D simulations of oscillations in prominences}
\authorrunning{A. Adrover-González \& J. Terradas}     
\maketitle      
        
\section{Introduction}
Solar prominence (or filament) oscillations have been the subject of study for decades. The first detection of a winking filament dates back to 1930 \citep[see][]{1966ZA.....63...78H}. Since then, diverse researches have aimed at establishing theoretical models that explain the examined oscillations. Observations together with the theoretical models have allowed us to expand our knowledge of solar prominences and have facilitated, for example, the estimation of physical parameters that are difficult to measure, such as the minimum magnetic field strength that supports these plasma structures, or the radius of the curvature of magnetic dips ($R$) \citep{2014IAUS..300...30B, 2014ApJ...785...79L}.

Because observational techniques have improved, the oscillations are analysed in more detail so that we can classify this phenomenon as large amplitude oscillations (LAO) or small amplitude oscillations (SAO) \citep[see][]{2002SoPh..206...45O}. The first group is characterised by velocity amplitudes of about $20~km~s^{-1}$ , and they are associated with an energetic event, such as solar flares \citep{2003ApJ...584L.103J,2012ApJ...760L..10L}, Moreton waves \citep{2002PASJ...54..481E,2008ApJ...685..629G}, Extreme ultraviolet Imaging Telescope (EIT) waves \citep{2004ApJ...608.1124O}, or coronal shock waves \citep{2011A&A...531A..53H,2014ApJ...795..130S} \citep[see][for a more detailed review of LAOs in prominences]{2009SSRv..149..283T}. In general, LAOs have been considered as a rare phenomenon, but \citet{2018ApJS..236...35L} have catalogued 196 filament oscillations in a period of six months (January to June 2014). Ninety of these events have been classified as LAOs. This indicates that LAOs are a frequent phenomenon. The first study of LAOs was carried out by \citet{1966AJ.....71..197R}, who investigated 11 winking filaments disturbed by flares. Winking filaments are periodic motions that are intermittently observed throughout different images shifted from the $H_{\alpha}$ line-centre by vertical oscillations; this causes a Doppler shift. The authors determined that the examined events oscillated at their own characteristic frequency. Then, using the results of \citet{1966AJ.....71..197R}, \citet{1966ZA.....63...78H} calculated the radial component of the filament magnetic field and the effective viscosity coefficients. He considered the movement as a damped harmonic oscillator embedded in a viscous fluid and in a uniform gravitational field, whose restoring force is the magnetic tension. \citet{1969SoPh....6...72K} studied horizontal transverse oscillations as a damped free oscillator with the magnetic tension as the restoring force, and radiation of acoustic waves as the attenuation mechanism. From these assumptions, they found that the oscillation period ($P$) for horizontal transverse oscillations is $P=4\pi LB^{-1}\sqrt{\pi \rho_p}$, where $2L$ is the length of the prominence, $B$ is the magnetic field strength, and $\rho_p$ is the prominence density.  \citet{2003ApJ...584L.103J} later proposed different mechanisms to explain the physics of an oscillatory motion along a filament observed on 2001 October 24. Firstly, they suggested as the driving mechanism the gravitational force when a concave upwards magnetic flux tube is considered as the main configuration of the cool filament thread. The second possibility that they proposed to explain the oscillations was a pressure imbalance that drives plasma motions along field lines. Finally, they suggested as a restoring force the magnetic tension for motions perpendicular to the local magnetic field. Furthermore, \citet{2007A&A...471..295V} detected oscillations along a filament on 2002 January 23 and proposed that the restoring force of the motion was caused by the magnetic pressure gradient along the filament axis, and \citet{2012ApJ...760L..10L}, who studied longitudinal oscillations in a solar filament on 2012 April 7, suggested that the restoring force was the coupling between the magnetic tension and gravity.

In contrast, SAOs are events with velocity amplitudes smaller than $10~km~s^{-1}$ , and they are not related to flare activity \citep{2012LRSP....9....2A}. These periodic motions are mostly understood as the oscillatory modes of magnetohydrodynamic (MHD) waves. To comprehend the nature of the oscillatory modes, the dispersion relation of ideal MHD equations has been derived for different models. \citet{1992A&A...256..264J,1992A&A...261..625J,1993A&A...277..225J} investigated in a set of three articles the normal oscillation modes of a 3D plasma slab embedded in an uniform longitudinal, transverse, and skewed magnetic field in the absence of gravity. The gravitational term was considered in \citet{1993ApJ...409..809O}. In a more complex configuration, \citet{2001A&A...379.1083D} examined the MHD waves of a single prominence fibril surrounded by coronal medium. From these articles a wide range of oscillatory modes were obtained and the oscillatory period of the prominences was estimated.   

The observational and theoretical studies of oscillations in filaments are corroborated by numerical simulations. Single 1D threads formed in magnetic flux tubes with a concave upwards shape where plasma oscillates have been studied by several authors. \citet{2012ApJ...750L...1L} studied large amplitude longitudinal oscillations in 47 independent threads. They found that the main restoring force is the projected gravity along the magnetic field, and that the periodicity of longitudinal oscillations depends on the radius of the curvature of the magnetic dip as

\begin{ceqn}
\begin{align}
P=2\pi\sqrt{R/g}~,
\label{eq.pendulum}
\end{align}
\end{ceqn}
where $g$ is the solar gravity acceleration. This model has been called the pendulum model. \citet{2012A&A...542A..52Z} used 1D radiative hydrodynamic simulations that almost reproduced the periodicity of longitudinal oscillations in threads with magnetic dips observed on 2007 February 8. \citet{2013A&A...554A.124Z} carried out a parametric survey to investigate the behaviour of longitudinal oscillations for different filament dimensions and triggering mechanisms. They found that  the swaying motion is not determined by the disturbance type and reasserted that the longitudinal oscillations follow the pendulum model. They also found that in addition to the gravity force, the pressure gradient contributes to the restoring force for short threads.

In addition to 1D simulations, 2D numerical studies of oscillations in prominences have also been investigated. Some of these works focused on studying the evolution of cold and dense plasma structures embedded in an isothermal stratified atmosphere permeated by a dipped arcade magnetic configuration. Based on this method, \citet{2013ApJ...778...49T} analysed the evolution of the density distribution together with other variables such as the involved forces or the velocity field during the relaxation process of the non-equilibrium initial configuration. Then, when the prominence was in equilibrium, they studied the behaviour of induced vertical, longitudinal, and Alfvén waves solving the linearised MHD equations. For validating purposes, \citet{2016ApJ...817..157L} reproduced non-linear numerical simulations of large amplitude longitudinal oscillations in a nearly equilibrium configuration. They found that the pendulum model is robust even in the range of short prominence field lines. In addition to this type of models, \citet{2016SoPh..291..429K}, based on the Pikelner model, studied the dynamics of a prominence perturbed by a pressure disturbance, and \citet{2019ApJ...884...74Z} analysed 2D non-adiabatic MHD simulations of longitudinal oscillations.
        
To reproduce more realistic situations, more complex 3D MHD numerical simulations are required. In the literature, rare 3D studies have analysed prominence oscillations. \citet{2016ApJ...820..125T} and \citet{2018ApJ...856..179Z} carried out simulations of prominences embedded in magnetic flux ropes. While \citet{2016ApJ...820..125T} focused their study on the evolution of the density profile for different resolutions and magnetic twist during the relaxation process and in the development of Kelvin-Helmholtz instabilities, \citet{2018ApJ...856..179Z} investigated the dynamics of longitudinal, vertical, and horizontal transverse oscillations triggered by velocity disturbances. The magnetic flux rope geometry is the most widely accepted filament model because it aligns the longitudinal filament axis along the magnetic field lines. However, the magnetic arcade configuration with a curtain-like prominence have not yet been used in 3D numerical simulations of oscillations in prominences.

An important characteristic of the reported events is that in most cases, the amplitude of the oscillations attenuates. To explain the observed damping, different theoretical mechanisms have been suggested, for example thermal damping such as radiative cooling \citep{2001A&A...378..635T} and heat conduction \citep{2007A&A...471.1023S}, ion-neutral collisions in partially ionised plasmas \citep{2007A&A...461..731F}, mass flows \citep{2008ApJ...684..725S}, wave leakage \citep{1999A&A...345.1038S}, or resonant absorption \citep{2002ESASP.506..629G} \citep[see the review of][]{2018LRSP...15....3A}. The numerical studies reviewed in this section made an effort to clarify the nature of the attenuation. \citet{2012ApJ...750L...1L} proposed that the main damping mechanism for their model is the accretion of mass onto the threads during prominence formation. \citet{2012A&A...542A..52Z} suggested that the non-adiabatic terms are responsible for the attenuation. They measured the importance of radiative loss and thermal conduction and concluded that in their thread model, the first term is responsible for the damping time ($t_d$). \citet{2019ApJ...884...74Z} studied the non-adiabatic effects in the attenuation of longitudinal oscillations in detail. They found that these processes are the primary agent that dissipates the oscillations, but wave leakage also plays an important role in dissipating the kinetic energy. \citet{2016ApJ...820..125T} have shown in their model that the attenuation of the vertical transverse oscillations is produced by resonant absorption, although Kelvin-Helmholtz instabilities could also provide a mechanism to attenuate the prominences. Despite these attempts to establish attenuation mechanisms, numerical studies caution that part of the damping may be due to computational causes such as numerical viscosity or dissipation. The numerical damping complicates the study of the attenuation mechanisms, and for this reason, many studies disregard the attenuation problem. While for longitudinal oscillations the non-adiabatic processes have been analysed in many studies, the resonant absorption in 3D simulations of transverse oscillations needs a deeper study.   
        
Longitudinal oscillations in solar prominences have been studied numerically by many authors, especially in 1D and 2D, and the pendulum model has been corroborated for different geometries. However, a parametric survey of longitudinal oscillations in 3D simulations is required. On the other hand, different models have been used to compare analytical expressions to the transverse simulated periods. For example, \citet{2013ApJ...778...49T} and \citet{2018ApJ...856..179Z}, in their simulations of vertical oscillations, used the slab and string model of \citet{1992A&A...261..625J}, respectively, and obtained large differences between the analytical and numerical results. Both studies obtained better correlations with the results of \citet{2001A&A...379.1083D}, however. This paper performs a 3D numerical study of oscillations in a curtain-shaped prominence permeated by a magnetic arcade that supports the dense and cool plasma against gravity. The magnetic configuration is based on the model proposed by \citet{1957ZA.....43...36K}. Unlike other works, in which the prominence mass is suspended above the photosphere, we anchored the plasma to the bottom boundary of the computational domain, which in our model represents the photosphere. Large-scale filament properties show that the cool plasma of many prominences joins the chromosphere,  and possibly the photosphere, through the filament barbs \citep{2014LRSP...11....1P}. Extensive filaments with a few points of contact with the surface (Hedgerow prominences) are the most common type \citep{1998assu.book.....Z,2015ASSL..415...31E}. In addition, \citet{2014ApJ...795..130S} observed transverse oscillations above the solar limb and suggested that the body oscillated as a rigid body with one end anchored on the Sun. However, this structure has not been investigated numerically so far. For these reasons, we considered that our model is a good approximation to study this rich prominence type. The aim is to compute the period and damping time for vertical and longitudinal/transverse horizontal motions triggered by velocity disturbances for different mass distributions. The primary initial configuration is described in Sect.~\ref{sec:configuration} and the numerical aspects are explained in Sect.~\ref{sec:numerical}. The computed results are displayed in Sect.~\ref{sec:results}. Discussion and conclusions are summarised in Sect.~\ref{sec:discussion}.

\section{Initial reference set-up}
\label{sec:configuration}
To numerically study the oscillations in prominences, an isothermal stratified background atmosphere has been configured as in \citet{2015ApJ...799...94T}. The coronal density profile is expressed as $\rho_c=\rho_0e^{-z/\Lambda}$ , where $\rho_0$ is the density of the reference level located at the bottom boundary of the computational box $z=0$. Unlike the works of \citet{2018ApJ...856..179Z} and \citet{2019ApJ...884...74Z}, who inserted a temperature transition between the photosphere and the corona, in our model we consider that the level $z=0$ represents the photosphere and that it is connected to the corona through the magnetic field \citep{2013ApJ...778...49T,2015ApJ...799...94T}.
\begin{figure} 
        \includegraphics[clip, width=8.8cm]{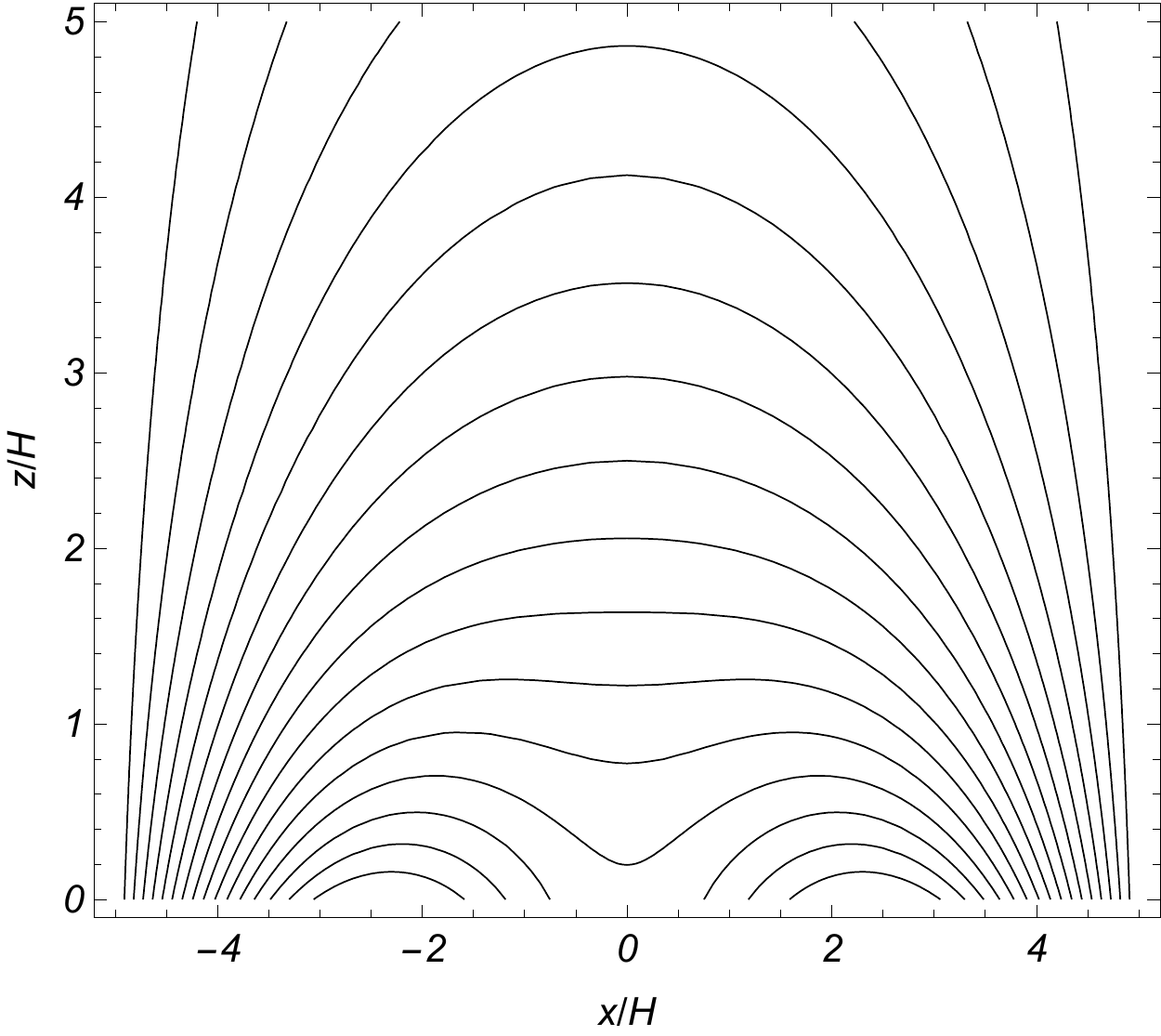}
        \caption{Magnetic field lines in the $xz$-plane. In this plot $B_0=10~G$, $k_1=\pi/2L$, $k_2=3k_1$, $L=5~H$, $z_0=-0.2~H$, and $l_1/k_1=1$. The $y$-component of the magnetic field is null. The maximum initial Alfvén speed is $v_{A0}=17.7~c_{s0}$.} 
        \label{fig:streamlines} 
\end{figure}
$\Lambda=c_{s0}^2/\gamma g$ is the density scale height. The velocity terms are normalised to the sound speed $c_{s0}=\sqrt{\gamma p_0/\rho_0}=166~km~s^{-1}$ for a coronal temperature of $1~MK$. The gravity acceleration is $g=0.274~km~s^{-2}$ and the adiabatic index  $\gamma=5/3$. The reference length has been taken as $H=10,000~km$. The plasma is permeated by an arch-shaped magnetic field, composed of the superposition of two force-free magnetic arcades. This quadrupolar configuration adds a magnetic dip to the structure. The depth of the dip varies with height. In Fig.~\ref{fig:streamlines} some magnetic field lines are plotted in the $xz$-plane. At $x=0,$ the radius of the curvature takes low positive values in the bottom part of the structure and grows until $z=1.55~H,$ where the field line becomes flat. From this point, the dip is lost and the curvature becomes concave downwards. To avoid high values of the plasma-$\beta$ parameter ($\beta=2c_s^2/\gamma v_A^2$), which affects the character of the modes, the null point of the magnetic force has been located out of the numerical domain as in \citet{2016ApJ...817..157L}. Unlike \citet{2015ApJ...799...94T}, the magnetic configuration has no shear and the magnetic field is invariant in the $y$-direction. An increase in shear produces an increase in the horizontal component of the magnetic field, leading to a higher magnetic tension. It also changes the depth and width of the dips so that the magnetic support can be different between sheared and non-sheared structures. The study of oscillations in sheared arcades is beyond the scope of this work. To obtain the previously described magnetic configuration, the magnetic vector potential is
\begin{equation}
\begin{cases}
A_x=0~,\\
A_y=\frac{B_0}{k_1}\cos k_1xe^{-l_1(z-z_0)}-\frac{B_0}{k_2}\sin k_2xe^{ -l_2(z-z_0)}~,\\
A_z=0~.
\label{eq.magnetic}
\end{cases}
\end{equation}
For the reference simulation we took $B_0=10~G$. $k_1=\pi/2L$ and $k_2=3k_1$ are parameters related to the lateral extension of the arcades, $2L$ is the width of the whole structure, $l_1$ and $l_2$ are a measure of vertical magnetic scale height, and $z_0=-0.2~H$ is the position of the null point. The relation $l_1/k_1$ is related to the shear of the magnetic field lines. In this work the shear is null so that $l_1/k_1=1$. $l_2$ is set by the constraint $k_1^2-l_1^2=k_2^2-l_2^2$. The magnetic field is $\textbf{B}=\nabla\times\textbf{A}$.

The mass deposition is artificial and instantaneous because we did not study prominence formation here \citep[see][for a summary of prominence formation models]{2010SSRv..151..333M}. The deposition maintains the gas pressure at the same level. The density profile consists of a curtain-shaped mass distribution expressed as

\begin{figure}
\includegraphics[trim =55mm 5mm 80mm 25mm,clip, width=8.8cm]{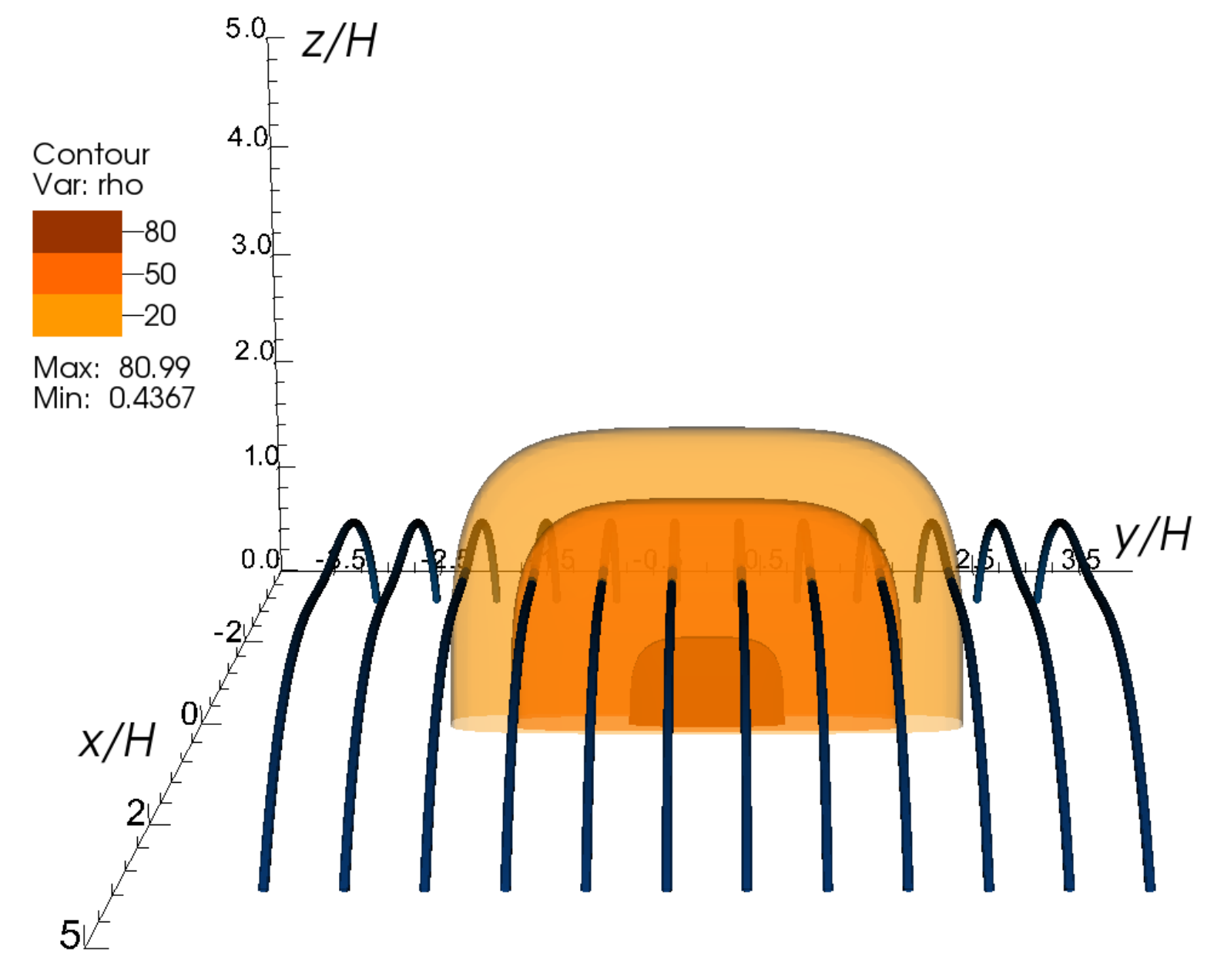} 
        \caption{Density profile together with some magnetic field lines at $t=0$. In this plot $n=4$, $w_x=0.3~H$, $w_y=2.2~H$, $w_z=2.6~H$, and $\rho_{p0}=80~\rho_0$.}
        \label{fig:init_conf}
\end{figure}

\begin{ceqn}
        \begin{align}
\rho_p=\rho_{p0}\exp(-2((x/w_x)^n+(y/w_y)^n+(z/w_z)^n)~.
\label{eq.density}
        \end{align}
\end{ceqn}
The term $n$ determines the width of the prominence-corona transition region (PCTR) (in this work, $n=4$ by default). The width, length, and height of the curtain are  $w_x=0.3~H$, $w_y=2.2~H,$ and $w_z=2.6~H,$ respectively, and the density contrast between the corona and the core of the prominence is $\rho_{p0}/\rho_0=80$. With these characteristics and by considering that the prominence density is $5.2\times10^{-2}~kg~km^{-3}$ , the total mass of the filament is $1.6\times10^{11}~kg$. The foot of the prominence is anchored to the base of the domain box, which represents the base of the corona, therefore the motion is subjected to this condition. The density profile is located centred on the computational box and the filament spine is  aligned with the $y$-coordinate so the magnetic field lines thread the prominence transversely. This configuration is not the most widely used by authors, who accept that in most cases the dense plasma is almost aligned along the magnetic field lines. Observations demonstrate that the angle between the magnetic field lines and the prominence spine is quite small and has a maximum around 25º \citep{1995ASSL..199.....T}. However, we decided to use our configuration for simplicity and as a preliminary step before sheared arcades are analysed. In Fig.~\ref{fig:init_conf} the described density profile is represented at $t=0$ with some selected magnetic field lines. 

\section{Numerical aspects}
\label{sec:numerical}

The ideal MHD equations we solved are
\begin{ceqn}
        \begin{align}
\dfrac{D\ln\rho}{Dt}=-\nabla\cdotp\textbf{u}~,
        \label{eq.continuity}
        \end{align}
\end{ceqn}
\begin{ceqn}
        \begin{align}
\frac{D\textbf{u}}{Dt}=-c_s^2\nabla\left( \frac{s}{c_p}+\ln\rho\right) -\nabla\Phi_{grav}+\frac{\textbf{j}\times\textbf{B}}{\rho}~,
        \label{eq.motion}
        \end{align}
\end{ceqn}
\begin{ceqn}
        \begin{align}
\frac{\partial\textbf{A}}{\partial t}=\textbf{u}\times\textbf{B}~,
        \label{eq.induction}
        \end{align}
\end{ceqn}
\begin{ceqn}
        \begin{align}
\rho T\frac{Ds}{Dt}=0~.
        \label{eq.entropy}
        \end{align}
\end{ceqn}
The set of governing equations are expressed in terms of the entropy rather than the more usual thermal energy or pressure variables. $c_p$ is the heat capacity at constant pressure and was assumed equal to one. $\Phi_{grav}$ is the gravity potential and $\textbf{j}$ is the electric current. The induction equation is solved in terms of the magnetic vector potential. $D/Dt=\partial/\partial t+\textbf{u}\cdotp\nabla$ is the advective derivative. The rest of characters have their normal meaning. 

The code we used to solve the MHD equations is the\textit{ \textup{Pencil Code}}\footnote{\href{http://pencil-code.nordita.org/}{http://pencil-code.nordita.org/}}. It is a publicly available model that uses sixth-order finite-difference schemes. High-order derivative schemes, such as the one used in the Pencil Code\textit{}, reduce the numerical dissipation, but to obtain consistency in numerical solutions, the code needs small amounts of diffusion to damp out the modes near the Nyquist frequency \citep{2003and..book..269B}. For this reason we introduced a simplified second-order hyperviscosity term \citep[see][]{2004PhRvE..70b6405H} and a shock viscosity term at the equation of motion. We did not use artificial viscosity in the other equations. The hyperviscosity is proportional to a diffusion coefficient. This coefficient should be as small as possible but sufficient to reduce the wiggles in the results. In addition, to avoid wiggles, we used fifth-order upwind derivatives for the advection terms $\textbf{u}\cdotp\nabla\ln\rho$ and $\textbf{u}\cdotp\nabla s$. The time step is a third-order Runge-Kutta scheme.

\textit{\textup{The Pencil Code}} uses three layers of ghost points to implement boundary conditions. In this work, closed boundary conditions were applied \citep[see][]{2016ApJ...820..125T}. This means that line-tying conditions were imposed at all the boundaries of the computational box. A line-tying boundary condition sets the three components of the velocity to 0, the normal component of the magnetic field is kept constant, and the density and entropy variables have their spatial derivatives be null. This condition imposed at the bottom boundary is crucial to mimic the purely reflecting conditions of the photosphere and to obtain the magnetic support. Because the Pencil Code uses the magnetic vector potential instead of the magnetic field, to fix the magnetic component perpendicular to the boundary, we applied to the three components of $\textbf{A}$ the condition of antisymmetry relative to the boundary value that vanishes the second derivative of $\textbf{A}$. The numerical domain consists of a box of $180\times144\times90$ mesh points. The dimensions of the box are  $100~Mm$ in the $x$-direction, $80~Mm$ in the $y$-direction, and $50~Mm$ in the vertical component. In this way, we imposed an equidistant grid of $556~km$. 

\section{Results}
\label{sec:results}
\subsection{Relaxation process}
\label{sec:relax}

\begin{figure}
        \includegraphics[trim = 5mm 40mm 5mm 40mm,clip,width=8.8cm]{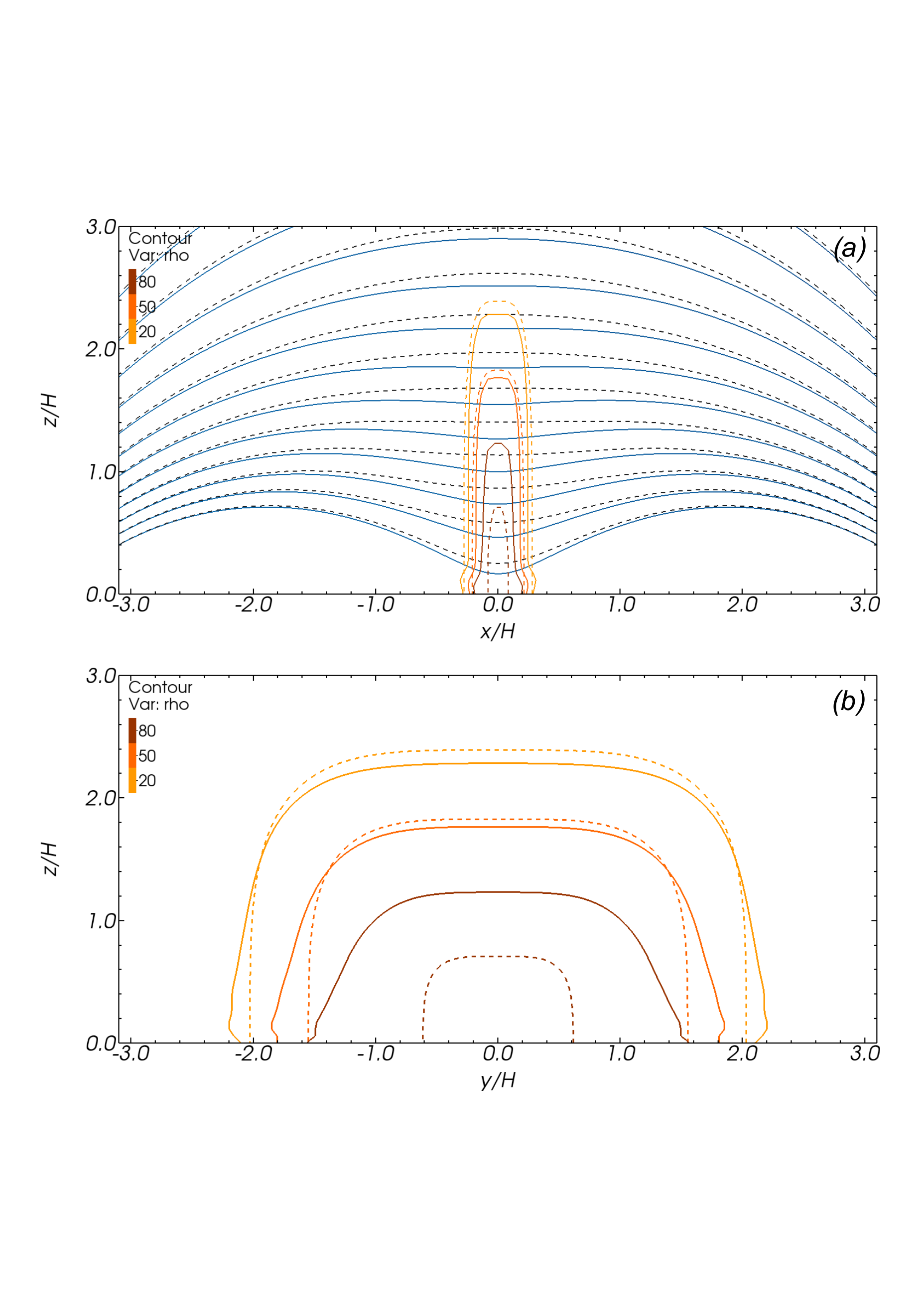}
        \caption{Density distribution of the final equilibrium state at $t=90~min.$ Top panel: Density isocontours and magnetic field lines in the plane $y=0$. Bottom panel: Density isocontours in the plane $x=0$. Dashed lines correspond to the initial reference set-up, and solid lines show the final equilibrium state at $t=90~min$.}
        \label{fig:final_state}
\end{figure}

As was mentioned in Sect.~\ref{sec:configuration}, the mass deposition is instantaneous. The prominence plasma breaks the initial equilibrium, and then starts to oscillate vertically. Initially, the force-free magnetic condition implies equilibrium between magnetic pressure and magnetic tension, and gas pressure force counteracts the gravitational force, so that the entire system is stable. However, when the prominence is introduced, the system loses its balance and the gravity force pushes the structure downwards. Immediately, the plasma deforms the field lines, increasing the depth of the magnetic dip. In panel $(a)$ of Fig.~\ref{fig:final_state} we show the final state of the field lines (blue solid curves) compared with the initial state (black dashed curves). This distortion implies an increase in the $z$-component of the magnetic tension (and to a lesser extent, the magnetic pressure) in such a way that it counteracts the movement, restoring the prominence upwards. After an evolution of approximately 80 minutes, the Lorentz force almost balances the gravity force. 

\begin{figure}
        \includegraphics[trim = 20mm 40mm 20mm 20mm,clip, width=8.8cm]{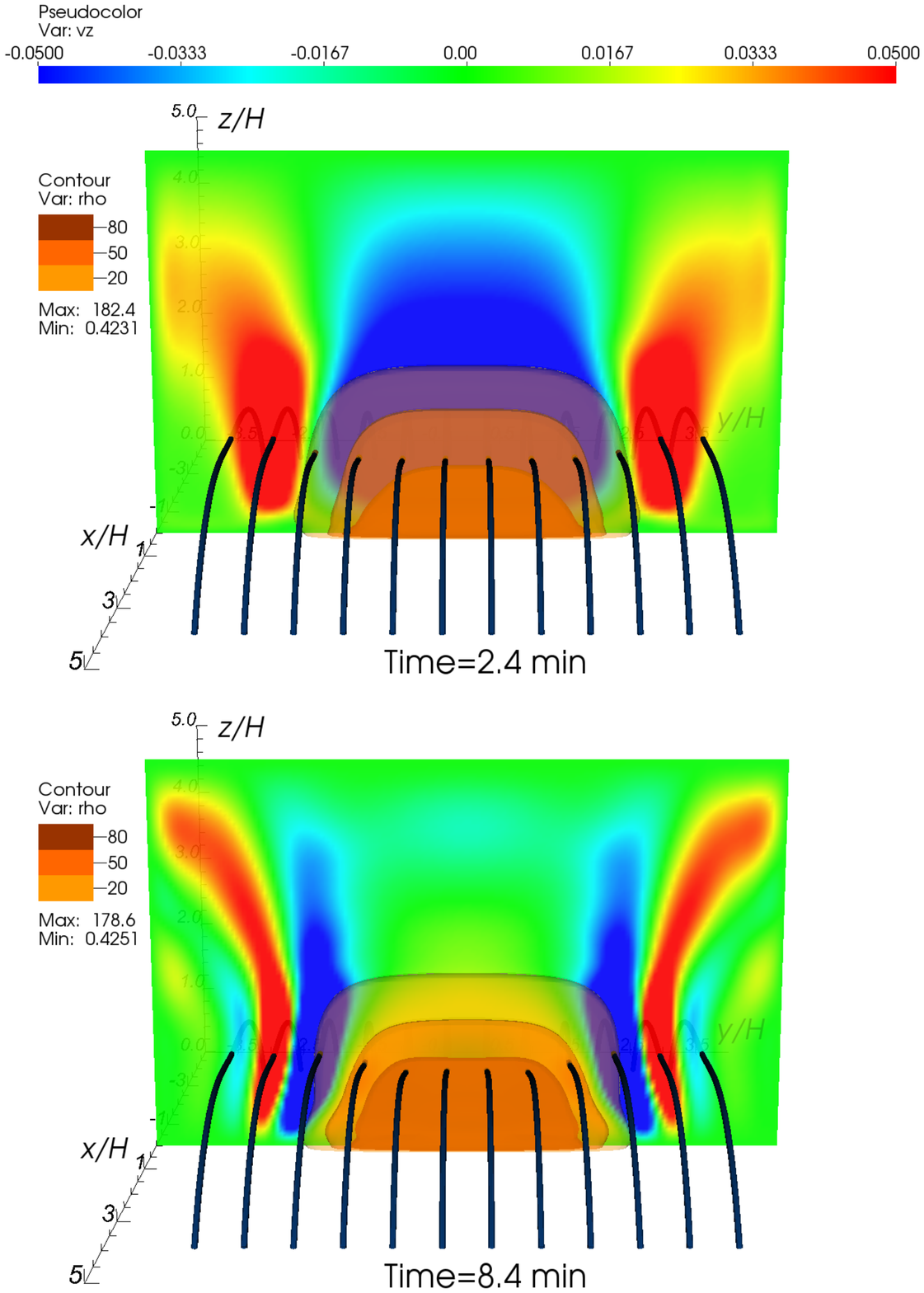}
        \caption{Time evolution of density, $v_z$, and magnetic field lines for the relaxation process. $v_z$ is represented in the $yz$-plane at $x=0$ as a 2D slice. The initial configuration for this case is shown in Fig.~\ref{fig:init_conf}.}
        \label{fig:uzslice}
\end{figure}

\begin{figure}
\includegraphics[trim = 0mm 0mm 0mm 0mm,clip,width=8.8cm]{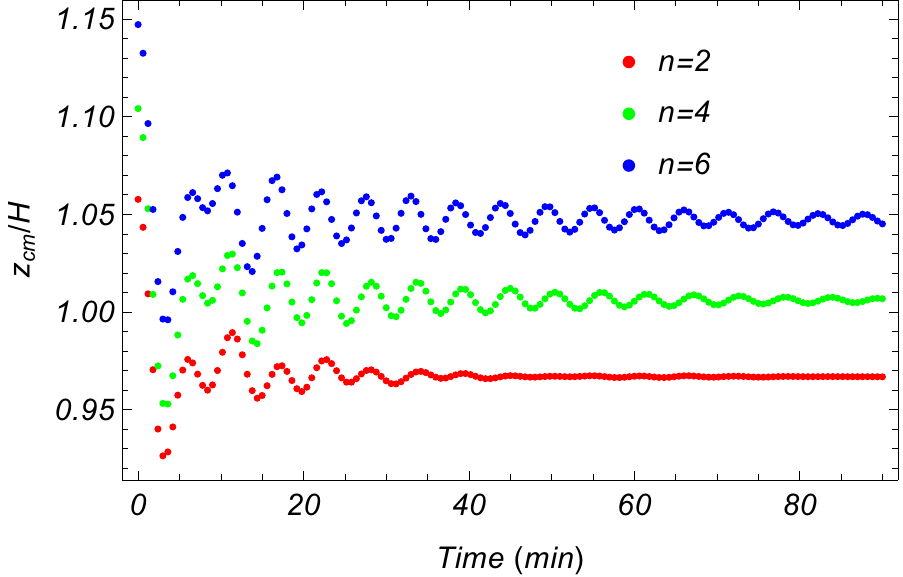}
        \caption{Time evolution of the z-component of the centre of mass for $n=2$ (red), $n=4$ (green), and $n=6$ (blue). The density average was computed for a delimited region ($x\in[-0.5~H,~0.5~H],~y\in[-3~H,~3~H],\text{and}~z\in[0,~3.5~H]$).}
        \label{fig:zcm}
\end{figure}

\begin{figure}
        \includegraphics[trim = 0mm 0mm 0mm 0mm,clip, width=8.8cm]{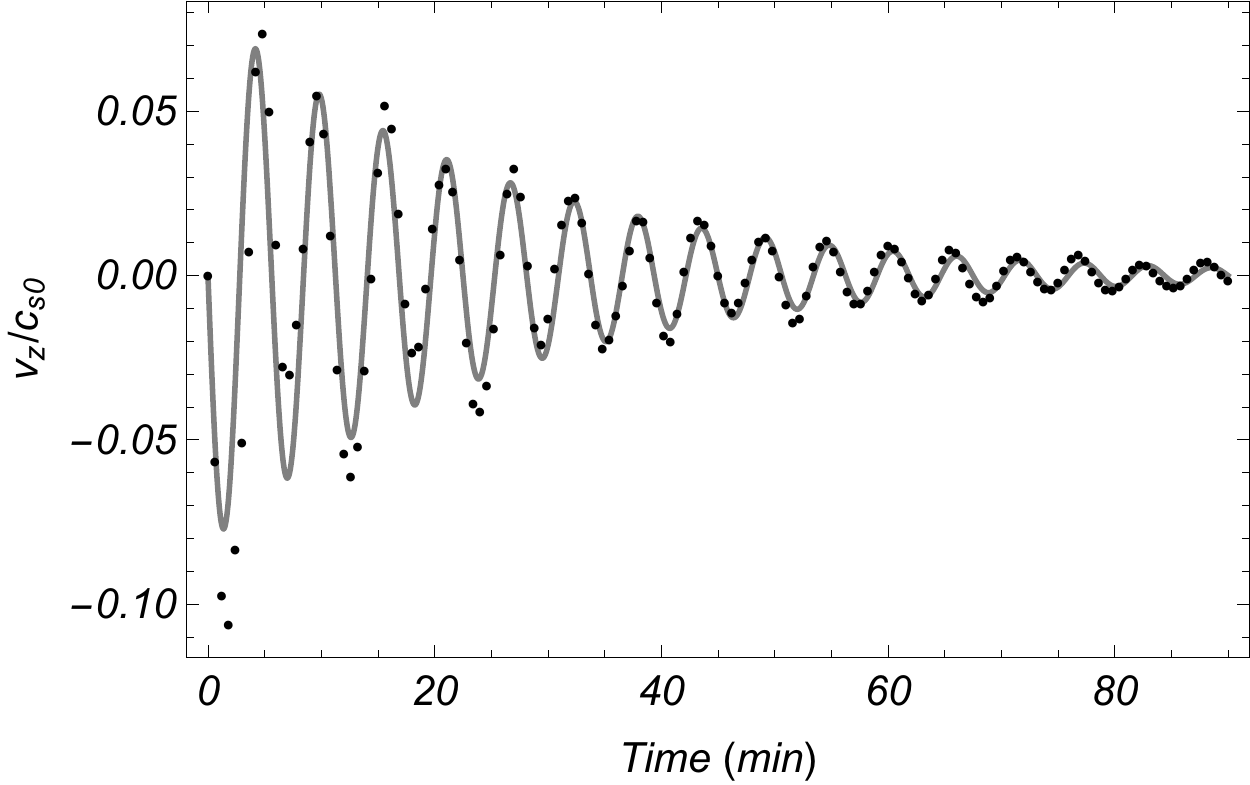}  
        \caption{$v_z$ as a function of time for the relaxation process at a fixed point ($0,0,\text{and }1.3~H$). Dots are the values obtained from the numerical simulation, and the continuous grey curve is the fitting result based on Eq.~(\ref{eq.adjust}).}
        \label{fig:uzpick}
\end{figure}

The total mass of the system remains practically constant, although Fig.~\ref{fig:uzslice} shows that the maximum density contrast between the reference level and the plasma slab increases. This involves the redistribution of the density profile. The density distribution of the final equilibrium state at $t=90~min$ is shown in the top and bottom panels of Fig.~\ref{fig:final_state}. In panel $(a)$ we observe that the prominence widens slightly in the $x$-direction near its base, where the plasma  gathers. To compensate for the accumulation of mass at the base, the filament narrows at the top and reduces its height. In panel $(b)$ we find that in the $y$-direction, the redistribution of mass is more noticeable, especially in the inner layers of the prominence, where as a result of the accumulation of plasma, the isocontour $\rho=80~\rho_{0}$ expands. The evolution of the $z$-component of the centre of mass is represented in Fig.~\ref{fig:zcm}. To suppress interactions with the background, the centroid was averaged from a delimited box that only included the prominence body. The green dotted line corresponds to the reference simulation, and, initially, the centre of mass is located approximately at $z=1.1~H$, however, it rapidly drops. After two cycles, the prominence oscillates around a new equilibrium height located at $z=1.01~H$. This fall agrees with the mass redistribution.

The evolution of the vertical velocity at the point $x=0$, $y=0$, $z=1.3~H$ is represented in Fig.~\ref{fig:uzpick} (dotted line). At the beginning of the simulation, the system is completely at rest, but the prominence body immediately drops, increasing the $z$-component of the velocity ($v_z$), which rapidly reaches its maximum value. Approximately at $t=3.5~min,$ the speed reaches positive values, which indicates that the movement begins to ascend towards its original position. This periodic motion is repeated approximately 16 times for about $90~min$. Fig.~\ref{fig:uzpick} shows that the amplitude of the oscillations decreases with time. The observations indicate that the movement describes an attenuated oscillation. We can fit an exponentially decayed sine curve with respect to time, expressed as 
\begin{ceqn}
        \begin{align}
        f=f_0\,\sin(\omega\,t)\,e^{-t/t_d}~,
        \label{eq.adjust}
        \end{align}
\end{ceqn}
where $f_0$ is the signal amplitude, $\omega$ the frequency, and $t_d$ is the damping time. For the reference simulation we obtained a period of $P =2\pi/\omega= 5.6~min$, a damping time of $t_d = 25.1~min,$ and a velocity amplitude of $f_0=-0.082~c_{s0}$. We represent the fitted curve in Fig.~\ref{fig:uzpick} as a solid line.    

\begin{figure}
\includegraphics[trim = 25mm 5mm 10mm 5mm,clip,width=8.8cm]{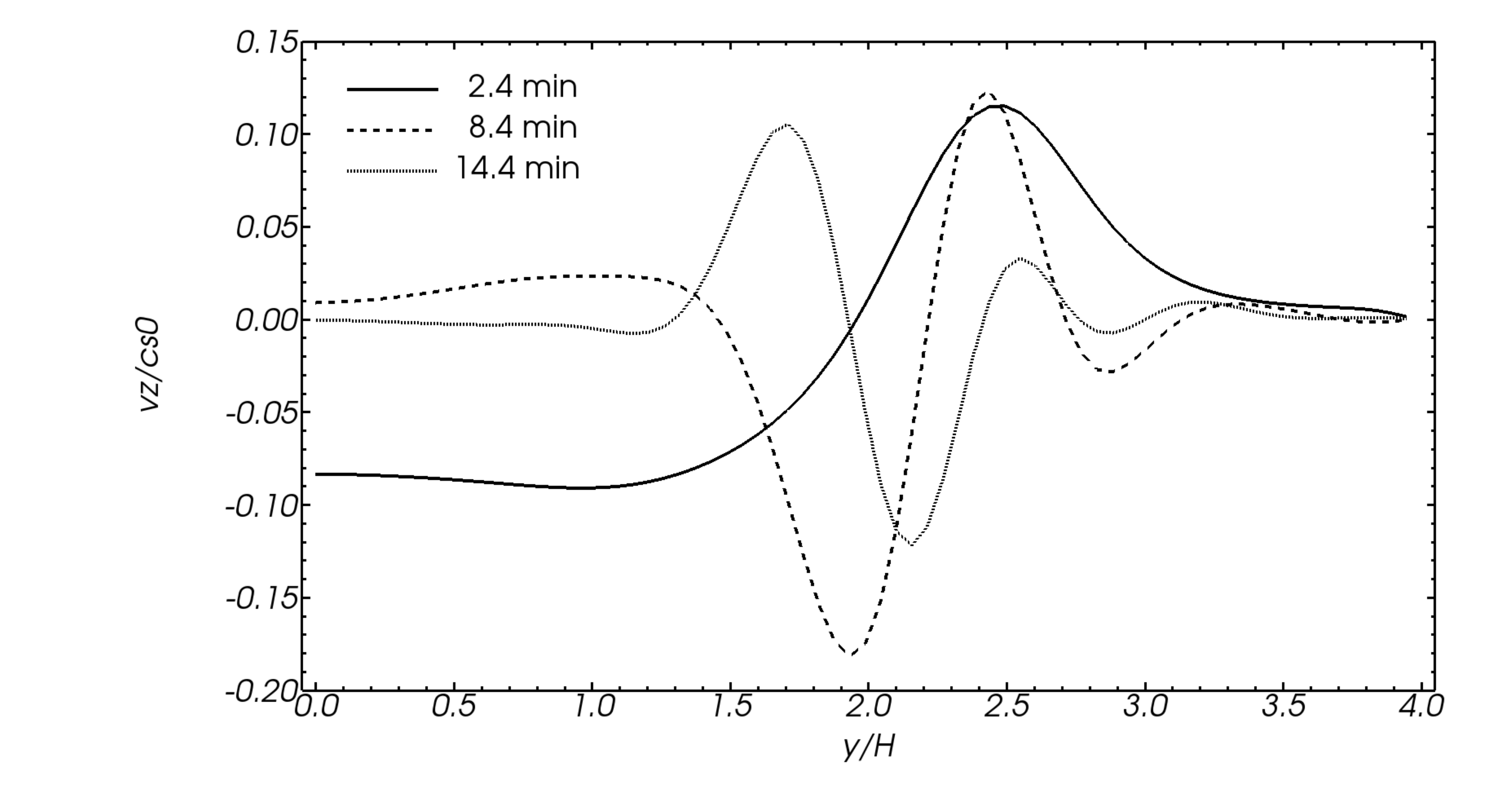}
        \caption{$v_z$ as a function of $y$ ($x=0$, $z=1.3H~$) at $t=2.4~min$ (solid line), $t=8.4~min$ (dashed line), and $t=14.4~min$ (dotted line).}
        \label{fig:vz_curves}
\end{figure}

As in \citet{2015ApJ...799...94T}, in Fig.~\ref{fig:uzslice} we represent the evolution of $v_z$ in a $yz$-slice at $x=0$. As we showed in Fig.~\ref{fig:uzpick}, the top panel of Fig.~\ref{fig:uzslice} shows that at the beginning of the simulation $v_z$ takes a negative value at the prominence body, describing a downwards movement. Nevertheless, the global evolution is more complex, and the system also develops vertical shear motions at the lateral edges of the prominence. In the first steps of the simulation, the system evolves a single wide lateral strip, but in the bottom panel of Fig.~\ref{fig:uzslice} we observe that at $t=8.4~min$ the number of bands with opposite speed has grown and their width has narrowed. This is clearer in Fig.~\ref{fig:vz_curves}, where we plot the time evolution of $v_z$ as a function of $y$-coordinate at $x=0$ and $z=1.3~H$. At $t=2.4~min$ (solid line) and $t=8.4~min$ (dashed line), we observe the same pattern as Fig.~\ref{fig:uzslice}, and at $t=14.4~min$ (dotted line), we obtain that these shear motions remain localised at the PCTR and that the typical spatial scales have decreased with time. This velocity pattern is generated as a result of resonant absorption, and it is the origin of part of the attenuation. \citet{2015ApJ...799...94T} also found this positive/negative velocity pattern in $v_z$ (see Fig.~3 of their work). They associated this behaviour with the process of mode conversion and phase mixing. Our results agree with the resonant absorption process because the energy conversion takes place at the inhomogeneous layer of the prominence, that is, at the sides of the structure in the transverse direction  with respect to the field lines and the direction of motion; in this case, the $y$-edges. This process is also characterised by the decrease with time of the typical spatial scales of the shear stripes.  

The studies of the resonant absorption as a mechanism for the damping of filament oscillations are usually based on simple models of magnetic flux tubes \citep[see e.g.][]{2008ApJ...682L.141A,2009ApJ...695L.166S}. In these works, the damping time of the fast kink waves caused by the resonant absorption process depends on the width of PCTR. To study the dependence of the attenuation on the PCTR, the relaxation process was simulated by modifying the density profile but keeping the total mass constant, so that for $n=2$, $\rho_{p0}=144~\rho_0$, and for $n=6$, $\rho_{p0}=63.7~\rho_0$. The prominence with the term $n=6$ has the steepest profile, and that of $n=2$ is the least steep, so that when $n$ increases, the density slope is stronger and leads to thinner transitions between the core and the corona. In Fig.~\ref{fig:zcm} the evolution of the centre of mass is plotted for the three cases. The evolutions of the three curves are similar each other, but we clearly see that for $n = 6$ (blue curve), the evolution has a softer attenuation and for $n = 2$ (red curve) we obtain the strongest damping. In agreement with \citet{2008ApJ...682L.141A}, we obtain that the wider the PCTR, the stronger the attenuation. We are interested in estimating the ratio of the damping time to the period. From the fitted curve of Fig.~\ref{fig:uzpick} we computed $t_d$ from the $v_z$ series at a localised point. However, in complex configurations we can detect different attenuation times at distinct points. For this reason, in Fig.~\ref{fig:zcm} we calculated the time evolution of the spatially averaged density. For the relaxation process, to calculate the damping time of the three signals of Fig.~\ref{fig:zcm}, we firstly applied an empirical mode decomposition \citep[see][]{2004ApJ...614..435T} to decompose the signal and select the first intrinsic mode function. In this way, we subtracted the trend of the centre of mass drop. After this, we obtained a damping per period of $t_d/P=1.9$ for $n=2$, $t_d/P=3.7$ for $n=4,$ and $t_d/P=5.9$ for $n=6$. 

\begin{figure}
        \includegraphics[trim = 5mm 5mm 0mm 5mm,clip, width=8.8cm]{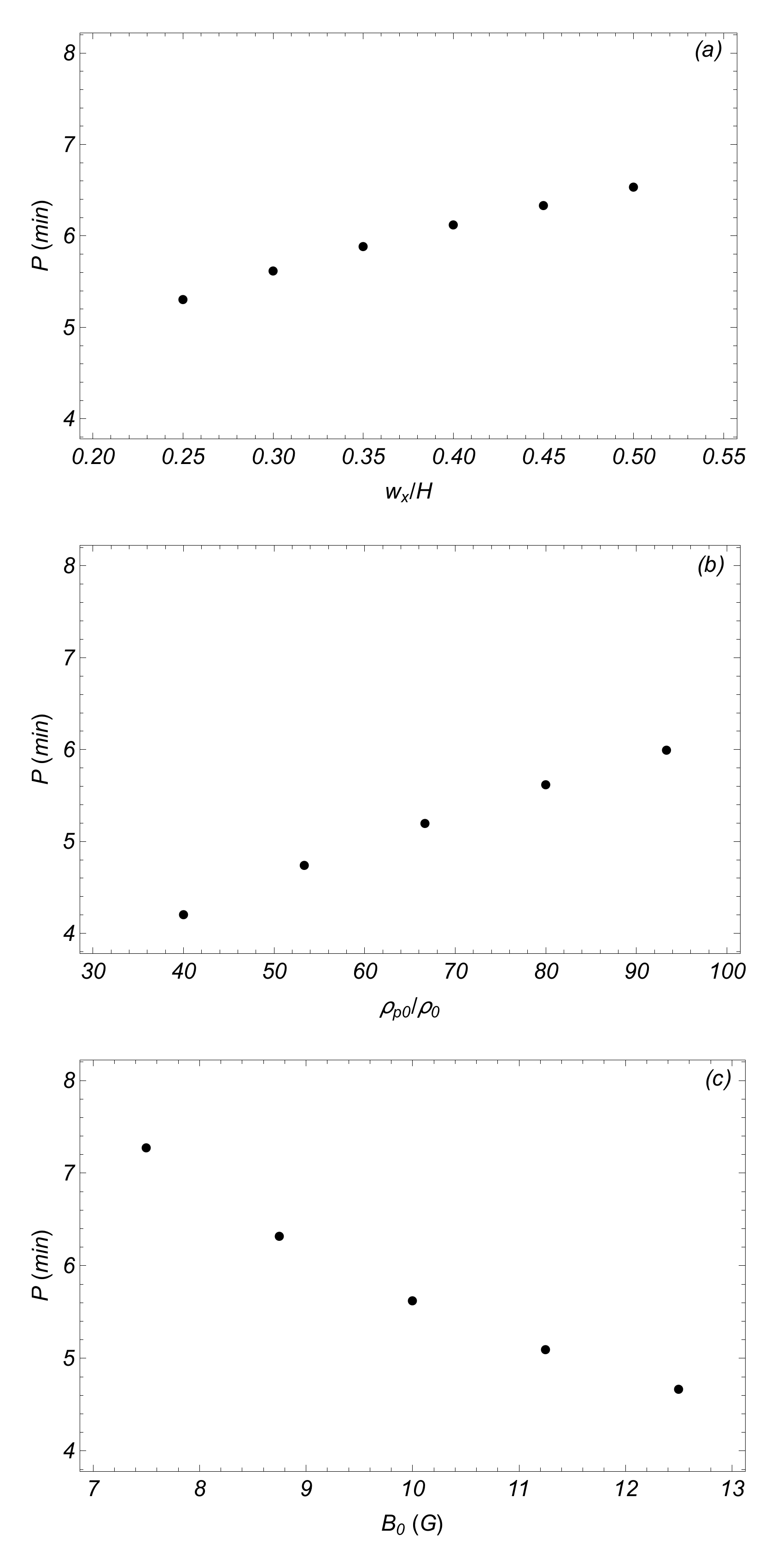}
        \caption{Scatter plots of the oscillation period during the relaxation process as a function of $w_x$ $(a)$, $\rho_{p0}$ $(b)$, and $B_0$ $(c)$. The values of $P$ were obtained from the fitted curve based on Eq.~(\ref{eq.adjust}) for the evolution of $v_z$ at a fixed point $(0,0,\text{and }1.3~H)$.}
        \label{fig:p_others}
\end{figure}

To extend the study of the relaxation process, we studied its evolution for a range of values of the filament width, the density contrast, and the magnetic field strength. The results are shown in Fig.~\ref{fig:p_others}. First we investigated the dependence of the period on $w_x$. We varied the width of the prominence from $0.25~H$ to $0.5~H,$ and for the other parameters, we used the reference values. When the width is doubled, the total mass is multiplied by a factor of two. To compute the periodicity of the oscillations, we fitted the results for the different time series of $v_z$ with Eq.~(\ref{eq.adjust}) following the same procedure as used in the reference process. The curtain with the widest width, and therefore with the greatest total mass, shows a higher initial $z$-velocity amplitude and a greater fall of the centroid into the new equilibrium point. Panel $(a)$ of Fig.~\ref{fig:p_others} indicates that the oscillation period varies with the prominence width. We obtained that the wider the prominence, the longer the period. 
 
We also studied the dependence of the periodicity on prominence mass. We varied $\rho_{p0}$ and kept the reference values of the other variables. Now the total mass of the parametric survey ranges from $7.9\times10^{10}~kg$ to $1.8\times10^{11}~kg$. The highest mass prominence suffers the greatest drop into the new equilibrium point and has the largest $z$-velocity amplitude. In panel $(b)$ of Fig.~\ref{fig:p_others} we show an increase in period with total mass. In panels $(a)$ and $(b)$ we observe an almost lineal dependence of the period on the other two parameters. The parameter values were  chosen so that the curtain with a width of $0.25~H$ and $0.35~H$ would have the same total mass as the curtain with $\rho_{p0}=66.67~\rho_{0}$ and $\rho_{p0}=93.33~\rho_{0}$ , respectively. Fig.~\ref{fig:p_others} shows that configurations with the same total mass have practically the same oscillation period.  

The last parameter we modified is the magnetic field strength. We varied $B_0$ so that the maximum Alfvén speed ranges from $13.3~c_{s0}$ to $22.1~c_{s0}$. \citet{2015ApJ...799...94T} analysed the dependence on $\beta$ of the relaxation process for a suspended prominence. They found that  for low $\beta$, the magnetic field of their configuration is able to keep the prominence suspended. However, for low magnetic field strength, the dense plasma falls down and finally forms a hedgerow or curtain-shaped prominence. These downward flows due to Rayleigh-Taylor instabilities complicate the dynamics of the system. Nevertheless, we avoided these flows by anchoring the density enhancement to the computational box. We obtained that for a weak magnetic field, the magnetic tension can hardly maintain the structure so that the material oscillates with a larger amplitude and at a lower equilibrium point. Panel $(c)$ of Fig.~\ref{fig:p_others} shows that the period decreases with the magnetic field strength. These results agree with \citet{2013ApJ...778...49T}, who found in their numerical study that for vertical oscillations, the period increases when the total mass is increased and decreases for weaker magnetic field strength. They also found that the period decreases slightly with the total length of the magnetic field line that crosses the prominence centre.

We analysed the attenuation of the vertical oscillations for the different cases. From the $v_z$ series we obtained that the damping per period $t_d/P$ does not vary significantly with width and ranges from 4.4 to 5.2. However, we found that the attenuation depends strongly on $\rho_{p0}$ and $B_0$. $t_d/P$ decreases with $\rho_{p0}$ from 7.9 to 3.7 and increases with $B_0$ from 2.2 to 7.6. When instead of analysing the $v_z$ series, we study the centroid, as has been previously explained, no important changes in periodicity or attenuation are observed. This means, as in \citet{2008ApJ...679.1611T}, that the global mode is dominant everywhere inside the prominence and that the damping time is basically the same everywhere. 

\subsection{Longitudinal oscillations}
\label{sec:longitudinal}

\begin{figure}
        \includegraphics[trim = 35mm 80mm 10mm 55mm,clip,  width=8.8cm]{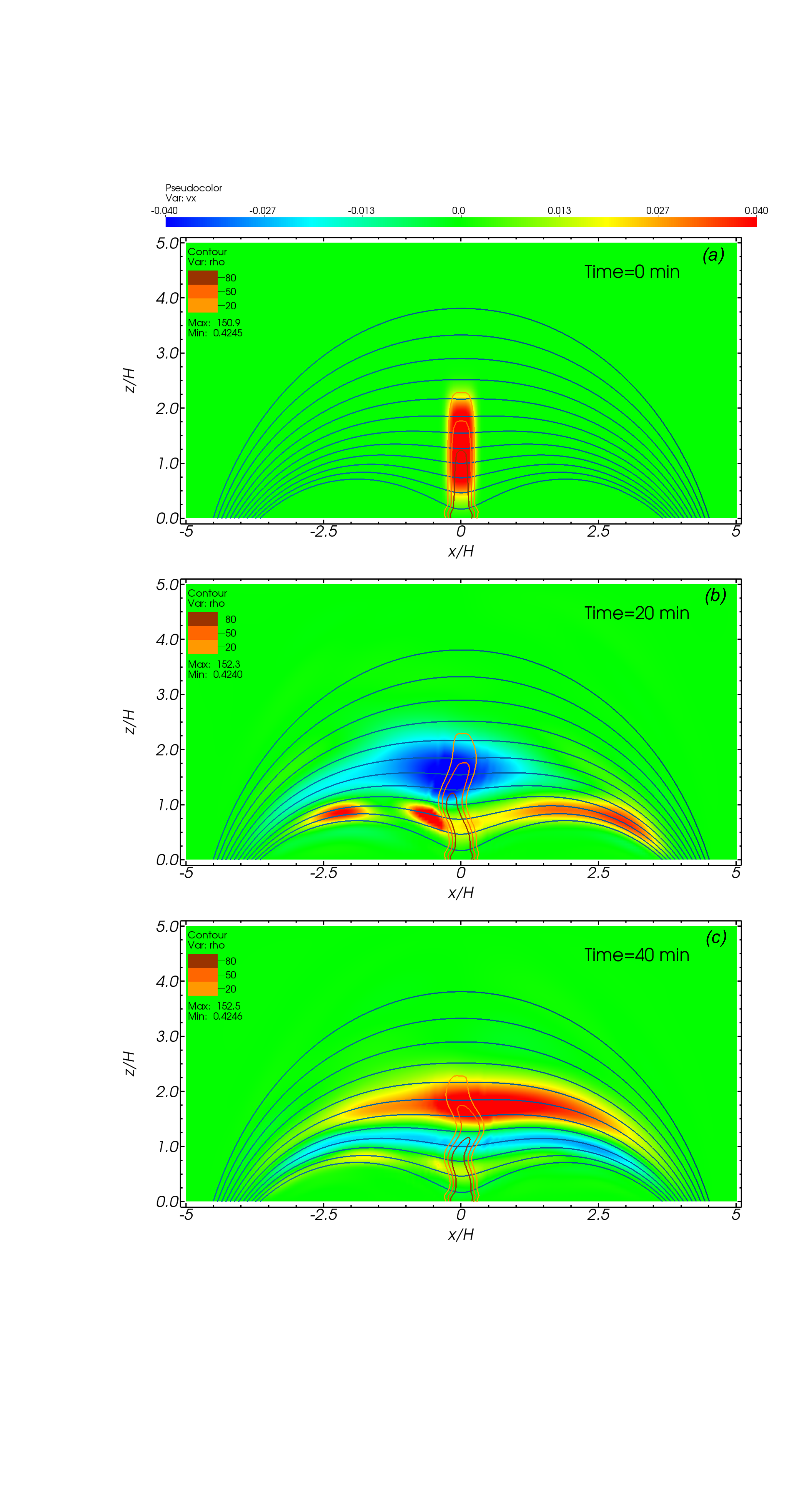}  
        \caption{Time evolution of $v_x$ at the $xz$-plane passing through $y=0$ for the evolution of longitudinal oscillations. The 12 blue curves correspond to some selected magnetic field lines. The orange isocontours represent the density profile. The initial parameters are the same as those in Fig.~\ref{fig:init_conf}.}
        \label{fig:vx_slice}
\end{figure}

When the dense plasma enters an almost stationary state, we introduce in the prominence body a velocity perturbation in the $x$-direction to trigger longitudinal oscillations. The perturbation is
\begin{ceqn}
        \begin{align}
        v_{xp}=v_{x0}\exp(-2((x/w_{vx})^4+(y/w_{vy})^4+((z-z_{v0})/w_{vz})^4))~.
        \label{eq.v_p}
        \end{align}
\end{ceqn}
We refer to longitudinal oscillations as the movement along the magnetic field lines. The spatial distribution of the velocity disturbance was defined in such a way that it produces a global motion of the prominence but maintains the filament foot fixed to the photosphere. The disturbance has a maximum speed $v_{x0}=0.05~c_{s0}$ located at the centre of the prominence at a height of $z_{v0}=1.3~H$, and its shape fits the density profile ($w_{vx}=0.3~H$, $w_{vy}=2.2~H$, and $w_{vz}=1.1~H$). In the top panel of Fig.~\ref{fig:vx_slice} the evolution of the $x$-component of the velocity ($v_x$) at the $xz$-slice passing through $y=0$ at the new equilibrium instant $t=0$ is plotted with some selected density contours and some magnetic field lines.

Initially, the prominence moves in the direction given by the disturbance. The displacement reaches a different maximum amplitude depending on the perturbation profile. The greatest displacement occurs in the centre of the structure where we located the maximum speed. After the oscillation reaches the maximum displacement, the backward movement restores the motion at a different instant for each height. The simulation shows that the restoring movement occurs earlier in the lower part of the structure than in the upper part. This fact causes the motion to describe a serpentine movement that can be seen in the middle and bottom panels of Fig.~\ref{fig:vx_slice}. Moreover, in the middle and bottom panel of Fig.~\ref{fig:vx_slice} we observe that $v_x$ displays different velocity stripes of opposite sign aligned with the magnetic field lines that cross the prominence body. These positive/negative bands indicate that the body oscillates out of phase. On the whole, the centroid oscillates with a period of $P=37~min$, a damping time of $t_d=47.6~min,$ and a displacement of $f_0=0.093~H$. 
\begin{figure}
\includegraphics[trim = 5mm 0mm 5mm 0mm,clip,  width=8.8cm]{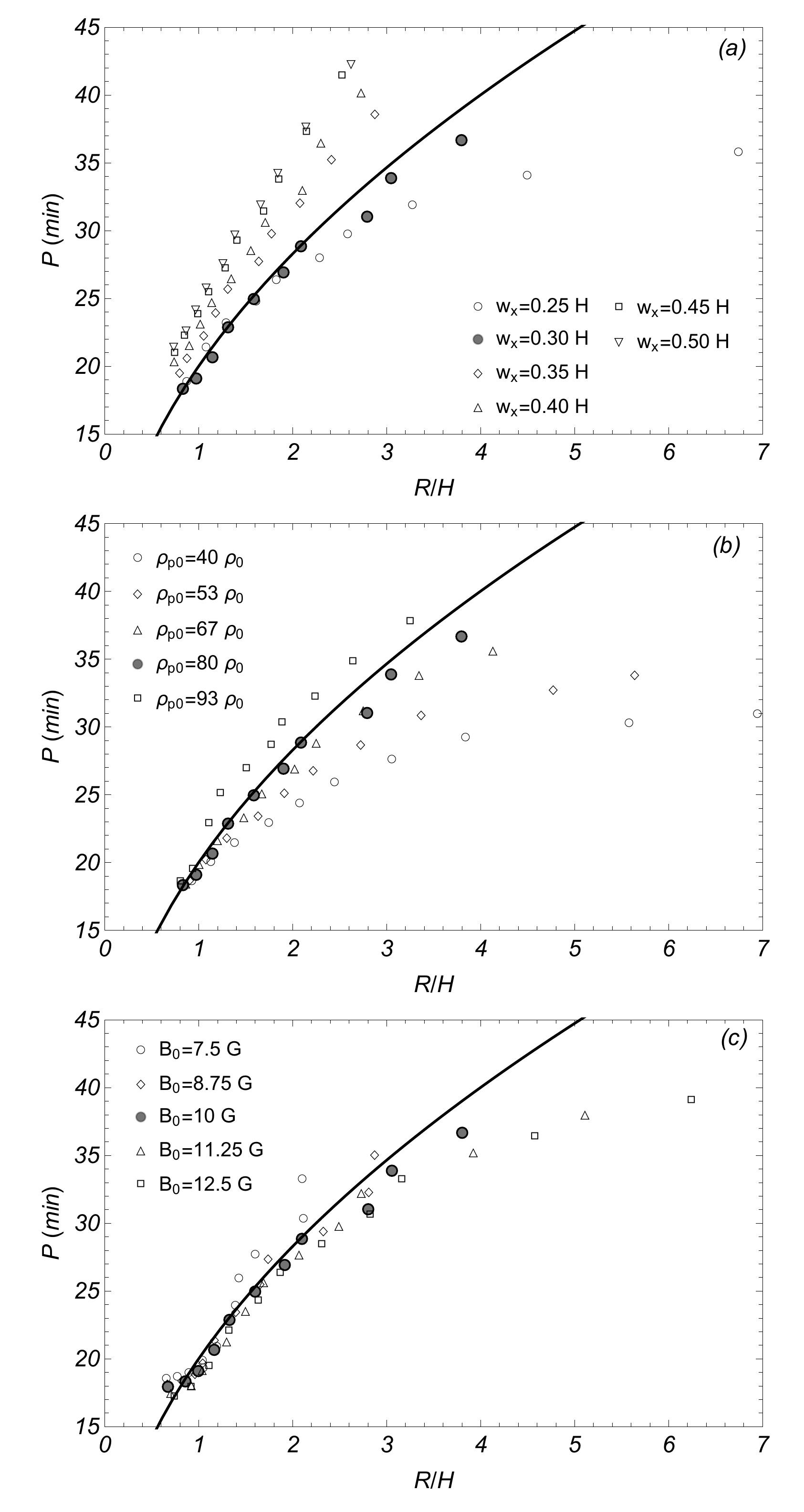}  
        \caption{Scatter plots of the oscillation periods for longitudinal oscillations as a function of the radius of the field line curvature. The top panel represents the different cases as a function of $w_x$, the middle panel as a function of $\rho_{p0}$, and the bottom panel as a function of $B_0$. The filled circles correspond to the reference simulation. The solid line represents Eq.~(\ref{eq.pendulum}).} 
        \label{fig:sl_alcases}
\end{figure}

To study the dependence of the period on height and the driving mechanism of the longitudinal oscillations, we compared the results with the  pendulum model. Based on \citet{2016ApJ...817..157L}, we selected 11 magnetic field lines located at the central $xz$-plane of our numerical domain. The position of the centre of mass along each magnetic line ($s_{cm}$) was calculated. The left foot position of the selected field lines ranges from $x=-4~H$ to $x=-3.65~H$ with an increase of $0.035~H$. The height of the selected lines at the centre of the prominence varies approximately from $z=0.3~H$ to $z=1.4~H$ for the reference simulation. At these heights all the selected field lines are concave upwards. We calculated the oscillation periods from the time evolution of $s_{cm}$ of each selected field line and represent them in Fig.~\ref{fig:sl_alcases} as a function of the radius of the field line curvature. To calculate $R$ we ignored its changes during the evolution of the oscillations, and we averaged the different values of the radius along a distance of $0.25~H$ from the prominence centre. The three panels of Fig.~\ref{fig:sl_alcases} show that the numerical results for the reference simulation (filled circles line) agree with the theoretical model expressed by Eq.~(\ref{eq.pendulum}) (solid line). 

In order to perform a parametric study, we introduced a $v_x$ disturbance in each of the different stationary states obtained in Sect.~\ref{sec:relax}, and analysed the periodicity of longitudinal oscillations as a function of $w_x$, $\rho_{p0}$, and $B_0$. In Fig.~\ref{fig:sl_alcases} we show the scatter plots of the oscillation periods as a function of $R$ for the different events. Panels $(a)$, $(b)$, and $(c)$ correspond to simulations of different $w_x$, $\rho_{p0}$ and $B_0$ , respectively. As we showed in Sect.~\ref{sec:relax}, the structures with a thinner density distribution, a lower density contrast, and a greater magnetic field strength suffer a smaller fall of their centre of mass during the relaxation process. Consequently, the prominence mass deforms the magnetic field lines less strongly, so that the radius of each field line is longer. As a result, in Fig.~\ref{fig:sl_alcases} we observe that the dotted lines for $w_x=0.25~H$ (panel $a$), $\rho_{p0}=40~\rho_0$ (panel $b$), and $B_0=12.5~G$ (panel $c$) reach longer radii for field lines with the same footpoint. In panel $(a)$ of Fig.~\ref{fig:sl_alcases} we show that the period increases with the prominence width. As a result, the computed results become different from those of the pendulum model. This result agrees with \citet{2013A&A...554A.124Z}. They reported a numerical study of a parametric survey of longitudinal oscillations and found that $P$ increases slightly with the width of the thread. The authors proposed that the dependence of the period on the prominence width arises because in shorter threads the gas pressure gradient increases, contributing to the restoring force and shortening the periodicity. In panel $(b)$ of Fig.~\ref{fig:sl_alcases} we also show  that the oscillation period increases with the density contrast of the structure. In agreement with this result, \citet{2013ApJ...778...49T} found that higher mass prominences have longer periods than low-mass prominences. They proposed that the cause of the increase in periodicity is the decrement of the sound speed when the mass is increased. In contrast, in panel $(c)$ of Fig.~\ref{fig:sl_alcases} the dispersion in the scatter plot is small, and this means that there is no dependence of the period on the magnetic field. In a low-beta regime, longitudinal oscillations are associated with the slow oscillatory modes whose characteristic speed is essentially $c_{s0}$ . The period therefore does not change significantly with $B_0$. In panels $(a)$ and $(b)$ of Fig.~\ref{fig:sl_alcases} we also show that the differences obtained with respect to the reference simulation and therefore with the pendulum model are greater for larger radii. These results indicate that in addition to the curvature radius, the prominence width and density contrast also modify the period of the longitudinal oscillations. This means taht the pendulum model is only a first approximation of longitudinal oscillations. This result agrees with \citet{2012ApJ...757...98L}. They determined that for larger radii the pressure force contributes to the restoring force so that the pressure-driven term introduces a correction into the pendulum model. In addition, the discrepancies with the pendulum model could be due to the calculation of the radii of the magnetic field line curvature. $R$ is not uniform along the streamline. The distance along the field lines we used to calculate the averaged radii can modify the results considerably. Furthermore, in this research, the variations in time that $R$ suffers due to the residual oscillations of the relaxation process were ignored.

To estimate the damping time per period, we computed these two parameters from the evolution of the $x$-component of the centroid. We obtained that $t_d/P$ decreases slightly with $\rho_{p0}$ and $B_0$, from 1.7 to 1.2 and from 1.8 to 1.1, respectively. $t_d/P$ does not vary significantly with $w_x$ , whose values range between 1.2 and 1.5 (these results do not change if we calculate the damping from the $v_x$ series). The values of $t_d/P$ are much more lower for longitudinal oscillations than for the relaxation process. This significant damping in contrast to the transverse motion was also observed in the simulations of \citet{2016ApJ...817..157L}, who associated the damping mechanism for longitudinal motions with a numerical viscosity that cancels the motion by stress. The variations of $t_d/P$ along with $\rho_{p0}$ and $B_0$ could be understood by the correlation between the attenuation and the oscillation amplitude. For the same speed disturbance, the higher the density, the greater the inserted kinetic energy, and therefore the greater the oscillation amplitude. From the simulations we also obtained a slight increase in amplitude with magnetic force strength and so the variation of the attenuation, but the amplitude does not change with respect to $w_x$. Another mechanism that can explain the attenuation of longitudinal oscillations is the wave leakage. \citet{2019ApJ...884...74Z} found in a 2D non-adiabatic filament simulations that when the ratio of gravity to Lorentz force is close to unity, longitudinal oscillations deform the magnetic field, generating the subsequent transverse waves, which propagate away from the filament body. In our model we did not find significant deformation of the field lines and the system does not develop waves perpendicular to the magnetic lines (not shown here). This means that no wave leakage is observed in the simulations. This is probably due to the low $\beta$ regime considered in this work.

\subsection{Horizontal transverse oscillations}
\label{sec:transverse}

\begin{figure} 
\includegraphics[trim = 30mm 65mm 40mm 45mm,clip,  width=8.8cm]{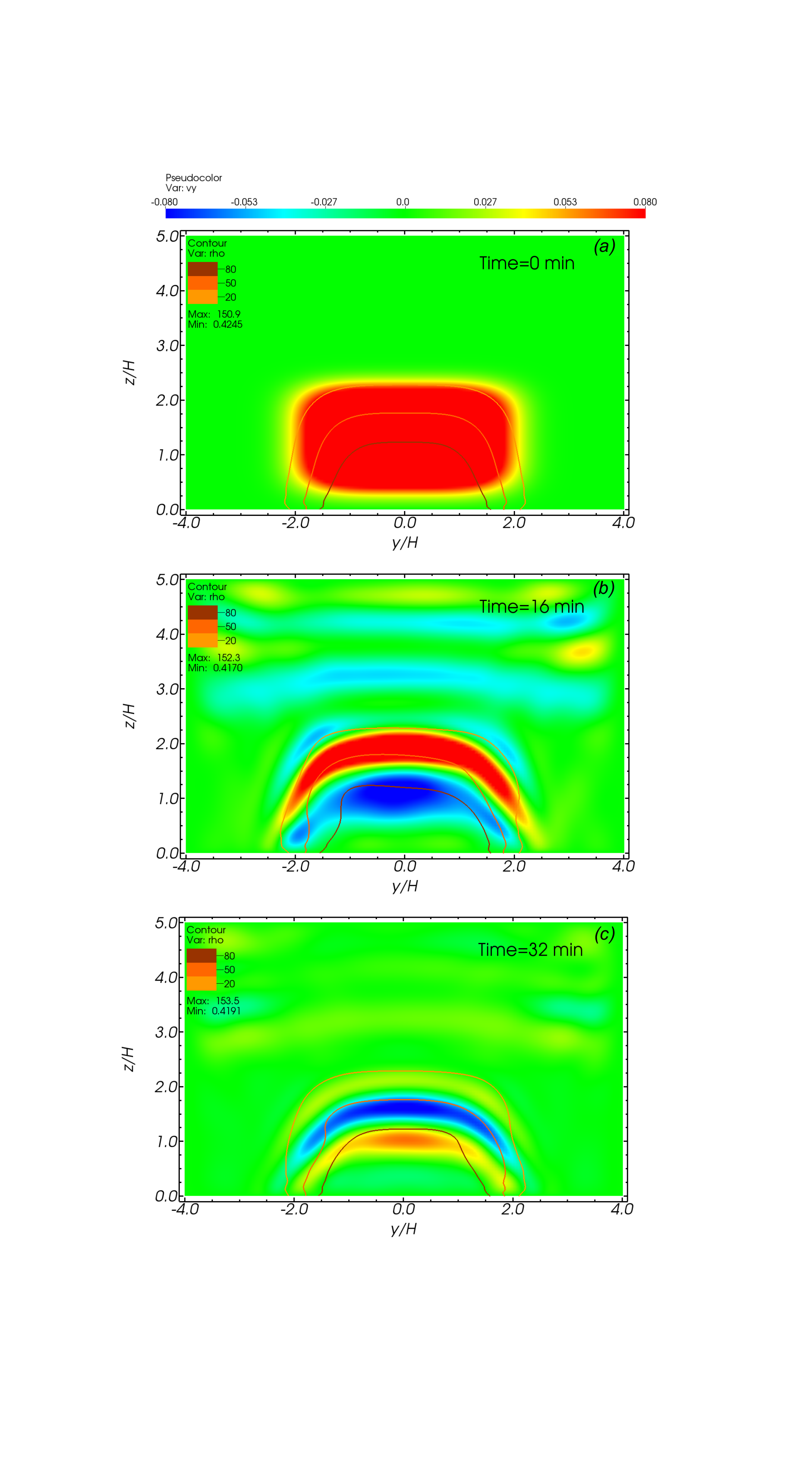}  
        \caption{Evolution of $v_y$ at the $yz$-plane passing through $x=0$ for horizontal transverse oscillations with the prominence density (orange isocontours). The initial parameters are the same as those in Fig.~\ref{fig:init_conf}.}
        \label{fig:vy_slice}
\end{figure}

To trigger horizontal transverse oscillations, we induced a velocity disturbance pointing in the $y$-direction. The spatial distribution of the velocity perturbation is the same as that introduced in Sect.~\ref{sec:longitudinal} for longitudinal oscillations, expressed as Eq.~\ref{eq.v_p}. In this case, the maximum speed located at the centre of the prominence at a height of $z=1.3~H$ is $v_{y0}=0.2~c_{s0}$. In the top panel of Fig.~\ref{fig:vy_slice} we show a $yz$-slice of the $y$-component of the velocity ($v_y$) passing through the central plane at $t=0$ together with some contours of the density profile. The plasma motion is somewhat complex. Initially, the inserted disturbance pushes the curtain, which is anchored at the base, tilting it towards the imposed direction. The displacement amplitude is greater in the internal parts of the filament where the velocity disturbance is stronger. Immediately, the plasma of the corona reacts to the movement in the opposite direction and tends to fill the displaced plasma. In addition, it is observed that the recovery movement starts at distinct times for different parts of the curtain. Regarding $v_y$, it is observed that during the first steps of the simulation the movement spreads as much throughout the prominence body as throughout the corona.
\begin{figure}
    \includegraphics[trim = 0mm 0mm 0mm 0mm,clip, width=8.8cm]{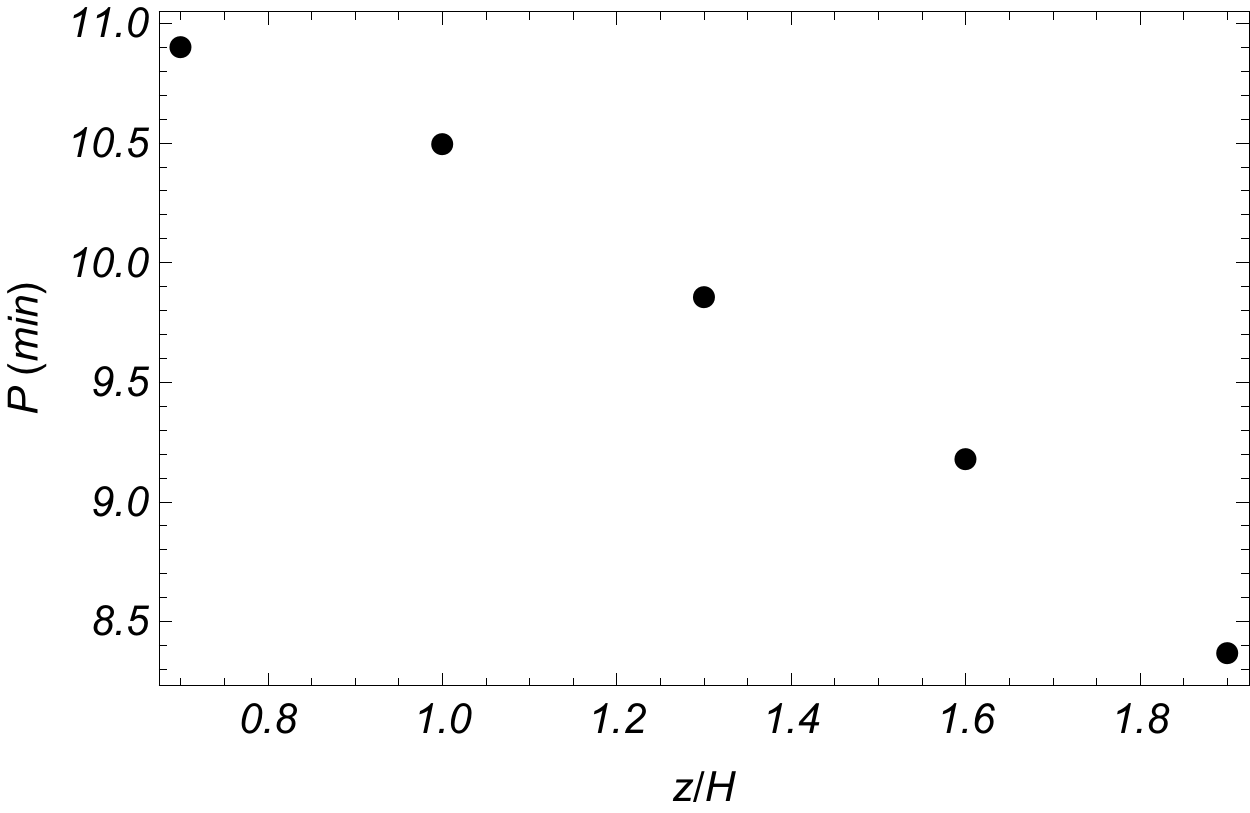}
        \caption{Scatter plot of the period as a function of height for horizontal transverse oscillations. The initial parameters are the same as those in Fig.~\ref{fig:init_conf}.}
        \label{fig:to_p_z}
\end{figure}
As time progresses, the oscillations are more localised at the prominence and its PCTR but attenuate at the corona. In the middle and bottom panel of Fig.~\ref{fig:vy_slice} we show that $v_y$ develops different stripes of opposite speed. These bands are localised in the prominence body but also in the background. The distribution of the velocity traces horizontal bands at the corona, but these stripes bend when they approach the prominence and take the shape of the density contour. These clear shear motions indicate differences in periodicity with height. The numerical results show that from the evolution of the $y$-component of the centre of mass, the oscillation period is $P=9.41~min$, the damping time is $t_d=11.9~min,$ and the displacement amplitude is $f_0=0.14~H$. As we expected, $P$ is shorter than that of longitudinal oscillations because in a low-$\beta$ regime slow modes have longer periods. We obtained that for transverse oscillations $P$ is longer than that of the relaxation process, however. As was mentioned, $P$ varies with height, therefore we plot in Fig.~\ref{fig:to_p_z} the oscillation period as a function of $z$. We computed $P$ by fitting the temporary evolution of $v_y$ for different heights with Eq.~(\ref{eq.adjust}) in which the sine was changed by a cosine. We obtained that the periodicity decreases with height. This result was also found by \citet{2018ApJ...856..179Z}. 

\begin{figure}
   \includegraphics[trim = 5mm 0mm 0mm 0mm,clip,width=8.8cm]{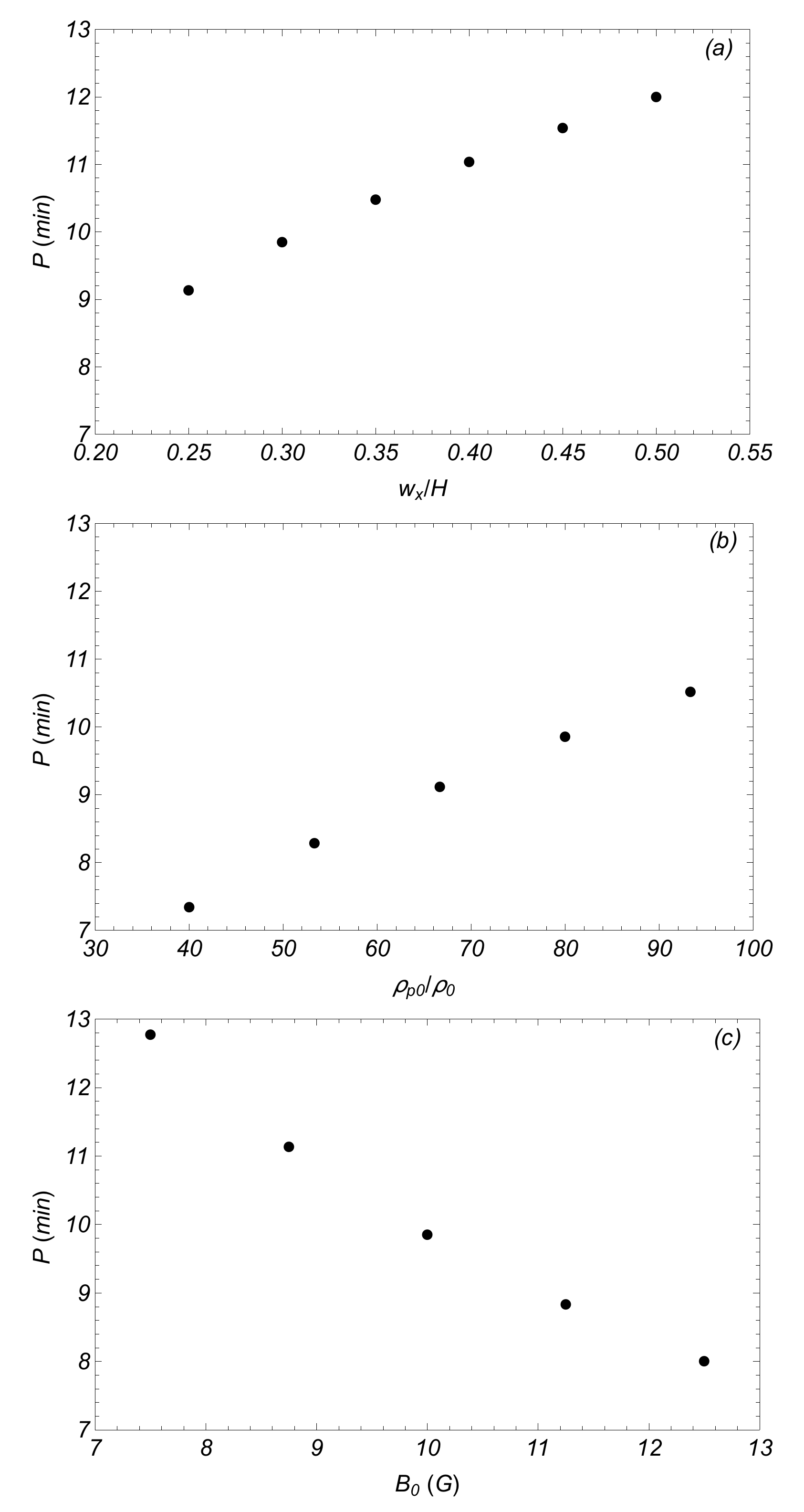}
        \caption{Same as Fig.~\ref{fig:uzpick}, but for horizontal transverse oscillations.}
        \label{fig:to_allcases}
\end{figure}

We again analysed the behaviour of horizontal transverse oscillations from the different events studied in Sect.~\ref{sec:relax}. In Fig.~\ref{fig:to_allcases} the scatter plots of $P$ are represented as a function of $w_x$, $\rho_{p0}$ and $B_0$. Panels $(a)$ and $(b)$ show that the periodicity increases with the prominence width and with the initial density contrast. However, panel $(c)$ indicates that $P$ decreases with the magnetic field strength. This behaviour coincides with the results observed in Fig.~\ref{fig:p_others} for the relaxation process. In both cases the decrease with the strength of the magnetic field is strong and the increase with density and prominence width is weaker. The fast mode of transverse oscillations is essentially a magnetic wave that is driven by magnetic forces; for this reason, we expected to find a stronger dependence of the period on magnetic field strength. 

The restoring force for the horizontal transverse oscillations is the magnetic tension force instead of gravity or pressure gradient. The different driving mechanisms for longitudinal and transverse oscillations reveal the different periodicity of the two oscillation modes.

While for vertical oscillations we obtained that $t_d/P$ varies with $\rho_{p0}$ and $B_0$, the damping in horizontal transverse oscillations does not change considerably with the considered parameters. When we compute $t_d/P$ from series of the $y$-component of the centroid, all events show values that range from 1.25 to 1.34. These values are higher (from 2.79 to 2.92) when the calculation is made from the series of $v_y$ at the point $x=0$, $y=0$, $z=1.3~H$. This means that we have different damping at different parts of the prominence. As was explained in Sect.~\ref{sec:relax}, the attenuation mechanism of the transverse modes is resonant absorption. For horizontal oscillations the absorption would occur in the upper part of the curtain, in its PCTR. However, as we showed in Fig.~\ref{fig:vy_slice}, the shear motions produced as a consequence of the attenuation are displaced towards the inside of the prominence. This behaviour is studied in more detail in Sect.~\ref{sec:modes}. The attenuation of transverse oscillations also depends on the PCTR. As we saw for the vertical oscillations, we obtained that the smoother the PCTR, the stronger the attenuation. We obtained that for $n=2$, $t_d/P=0.7$, $n=4$ $t_d/P=1.27$ and $n=6$ $t_d/P=1.87$, respectively.

\subsection{Continuum modes}
\label{sec:modes}

\begin{figure}
   \includegraphics[trim = 0mm 0mm 0mm 0mm,clip,width=8.8cm]{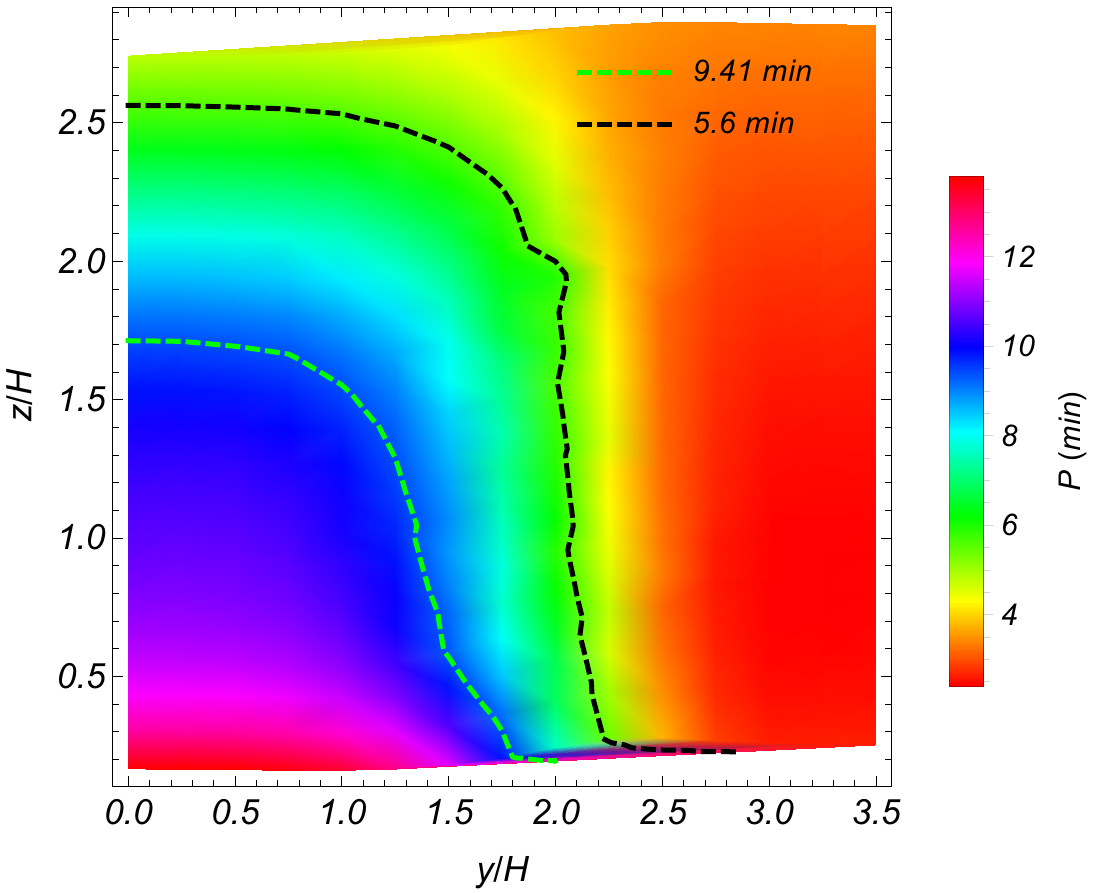}
        \caption{Alfvén continuum spectrum in terms of the periodicity as a function of the position of each field lines at $x=0$. Dashed black and green curves are the isocontours corresponding to the periods of the vertical and transverse oscillations, respectively, inferred from the time-dependent simulations.}
        \label{fig:amodes}
\end{figure}

\begin{figure}
   \includegraphics[trim = 40mm 30mm 40mm 30mm,clip, width=8.8cm]{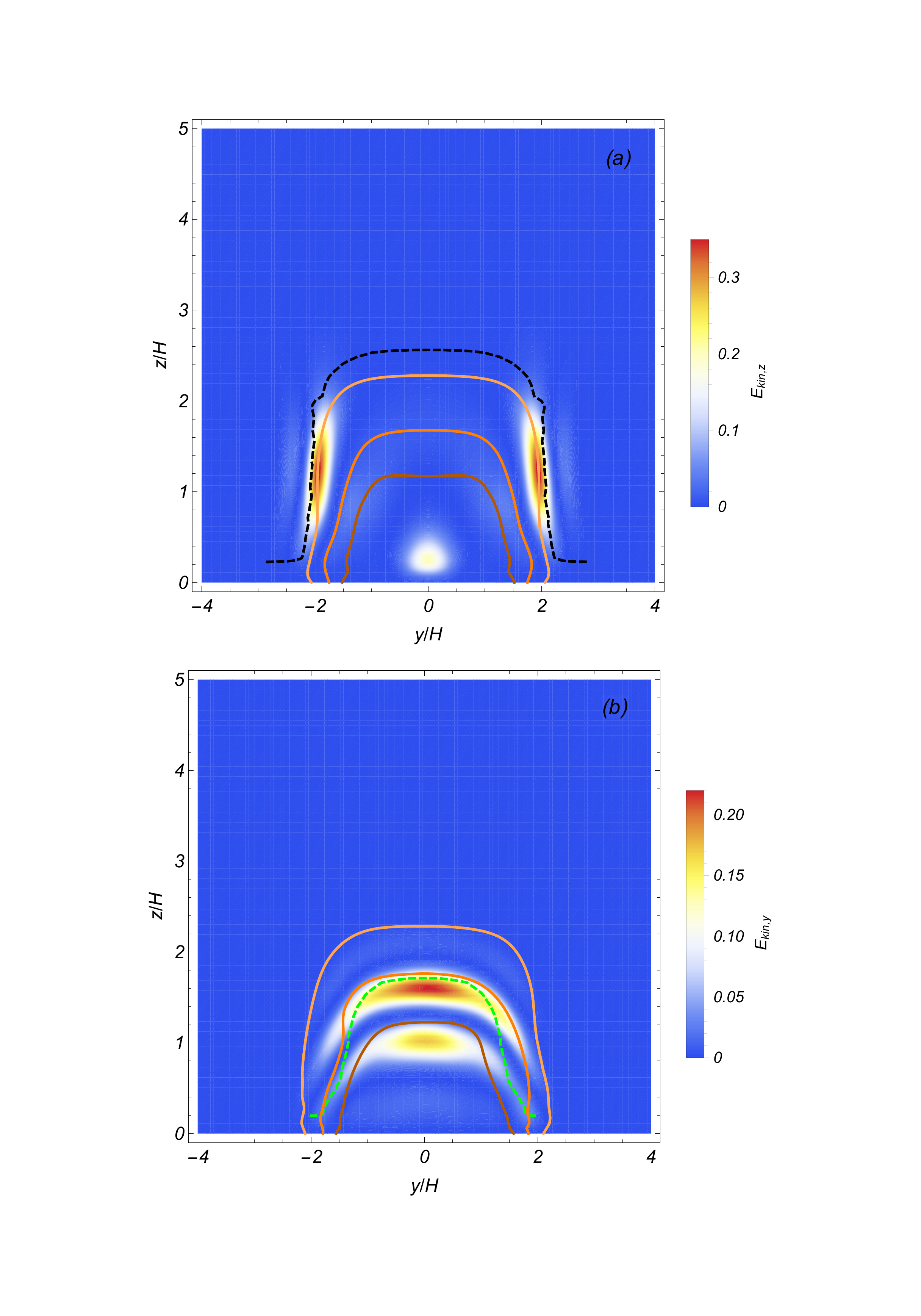}
        \caption{Top panel: Snapshot of the vertical kinetic energy and the density profile at $t=8.4~min$ with the vertical resonant surface (dashed black curve) obtained in Fig.~\ref{fig:amodes}. The orange density contours are the same as those in the bottom panel of Fig.~\ref{fig:uzslice}. Bottom panel: Snapshot of the $y$-component of the kinetic energy  and the density profile at $t=32~min$ with the horizontal resonant surface (dashed green curve) obtained in Fig.~\ref{fig:amodes}. The orange density contours are the same as those in the bottom panel of Fig.~\ref{fig:vy_slice}.}
        \label{fig:ekin_amodes}
\end{figure}

We have computed the Alfvén continuum modes of the MHD equations based on the expressions found in \citet{1985SoPh..102...51G}. They derived the continuous spectrum of a 2D magnetostatic equilibrium. Their study is invariant with respect to $y,$ therefore we considered our structure as $2D$ slices in the $xz$-plane. This is the main approximation because our prominence model is in 3D. The expression for the modes of the Alfvén continuum is written in \citet{2013ApJ...778...49T} (see their Eq.~(15)) in terms of the distance along the magnetic field lines. According to this equation, we can numerically compute the Alfvén frequency for different field lines by solving the eigenfrequency problem. To solve the equation, we need to extract the coordinates of each line and the variations of the magnetic field strength along the field line for the equilibrium state. We selected diverse magnetic field lines whose footpoints range from $-4.315~H$ to $-3.65~H$ in the $x$-direction and from $0$ to $3.5~H$ in the $y$-direction. The Alfvén spectrum is represented in Fig.~\ref{fig:amodes} in terms of the oscillatory period as a function of the position of each line at the central plane. The computation of the local Alfvén frequency is useful for understanding the resonant absorption process because, essentially, the resonant surface is the one where this frequency coincides with the frequency of the global mode inferred from the time-dependent simulations. In Fig.~\ref{fig:amodes} we show the contour of the global mode for the relaxation process (dashed black line) and for the horizontal transverse oscillation (dashed green line). For vertical oscillations, the resonant surface is more external than in the horizontal case. \citet{2008ApJ...679.1611T} found that the resonant surface can be located inside the structure, as we have obtained for the horizontal case. In Sect.~\ref{sec:relax} and Sect.~\ref{sec:transverse}, according to the velocity distribution of Fig.~\ref{fig:uzslice} and Fig.~\ref{fig:vy_slice}, we have estimated where the energy conversion takes place: at the $y$-edges of the prominence body for the relaxation case and at the sheared bands inside the prominence for the horizontal transverse oscillations. However, the location of the resonant absorption process is clearer when we study the kinetic energy of the system. The top panel of Fig.~\ref{fig:ekin_amodes} shows that the resonance surface of the vertical motion (dashed black curve) matches in space the location where the vertical kinetic energy increases due to the energy conversion. Only one snapshot of $E_{kin,z}$ at $t=8.4~min$ is represented, but the maximum of $E_{kin,z}$ is always localised around this point until when the oscillations attenuate. In the bottom panel of Fig.~\ref{fig:ekin_amodes} we show a snapshot at $t=32~min$ of the $y$-contribution of the kinetic energy for horizontal transverse oscillations. Again, the resonance layer almost coincides with the maximum energy of the system. The exact behaviour of 3D Alfvén resonances has been studied in \citet{2016ApJ...833..230W} and \citet{2017JGRA..122.3247E}. They found that in 3D an infinite number of possible resonant solutions exist within a resonant zone, where the energy is accumulated, on which a dominant contour stands out from different ridges. It is beyond the scope of the present work to calculate the 3D Alfvén resonances, and it is enough to mention that the 2D approach provides reasonable results.

\section{Discussion and conclusions}
\label{sec:discussion}

A 3D numerical study of transverse and longitudinal oscillations in filaments has been presented. A curtain-shaped prominence is anchored to the base of the corona so that the system does not develop downwards flows due to Rayleigh-Taylor instability, which complicates the analysis of the results of the oscillatory motion. In addition to the fixed foot of the structure, the behaviour of the system is subject to closed boundary conditions, meaning that we have imposed line-tying conditions at all sides of the computational box. This perfectly reflecting condition applied in the bottom plane is essential to support the structure. To study the attenuation mechanism for transverse oscillations, we analysed the velocity distribution of the motion, the dependency of the attenuation on the PCTR, and the continuum Alfvén modes. In agreement with our simulations, we proposed that the main damping mechanism is the resonant absorption. As the theoretical models predict, we obtained that the attenuation is stronger for wider PCTR. From the analysis of the Alfvén continuum modes, we obtained that the location of the resonant surface coincides with the concentration of the kinetic energy, confirming that the resonant absorption is the main mechanism for transverse oscillations. However, we obtained that $t_d/P$ ranges from 2.2 to 7.9 for the relaxation process of different scenarios, values that in principle are higher than expected from the observations. This means that in addition to the resonant absorption, another type\ of damping mechanism must be considered in our model. On the other hand, for transverse and longitudinal horizontal oscillations we obtained a stronger attenuation. One reason that can explain this significant damping for longitudinal oscillations could be changes in periodicity with height. The plasma of two adjacent magnetic field lines with different curvature radius oscillates at a different frequency, so that some shear may brake the movement. This type of sheared viscosity mechanism can also contribute to the attenuation of horizontal transverse oscillations. The shear motions inside the prominence due to the resonant absorption can brake the oscillation and increase the attenuation. Regardless, our model is relatively simple to study the observed damping, and other mechanisms such as non-adiabatic processes, mass flows, or partial ionisation effects must be considered. 

Regarding the periodicity, we reproduced oscillations with periods that range approximately from $4.2~min$ to $7.3~min$ for vertical oscillations, from $28.2~min$ to $42.3~min$ for longitudinal oscillations, and from $7.4~min$ to $12.0~min$ for horizontal transverse oscillations. These results are somewhat shorter than the observations. In \citet{2009SSRv..149..283T} the periods associated with vertical motions range from $15~min$ to $29~min$ and in \citet{2014ApJ...786..151S} from $11~min$ to $22~min$. In \citet{2018ApJS..236...35L} the periods for horizontal transverse oscillations range from $30~min$ to $80~min$ and for longitudinal events from $30~min$ to $110~min$. We recall that in the field of observations, in contrast to this work, longitudinal oscillations are motions along the filament axis. \citet{2010SSRv..151..333M} reported that the magnetic field rotates with respect to the filament axis by an angle in the range from 15 to 30 degrees. Then, \citet{2018ApJS..236...35L} in their catalogue considered as longitudinal oscillations those whose velocity vector forms an angle with the filament axis smaller than 40º. This suggests that a certain error is made in classifying the oscillation polarisation. Despite the short results obtained in $P$, for longitudinal oscillations we performed a reference simulation that agrees with the pendulum model. However, for values of $w_x$ and $\rho_{po}$ different from those of the reference simulation, the results move away from the theoretical expression. This means that another mechanism may drive the longitudinal motions, such as pressure force. For example, we obtained that the periodicity of the oscillations increases with the prominence width independently of the polarisation direction. This is an interesting result because it can be compared with diverse observed events.  \citet{2018ApJS..236...35L} assembled a valuable catalogue from which we can extract much information, especially from their statistics section. In agreement with our results, in Fig.~21 (panel $d$) of \citet{2018ApJS..236...35L} they show that $P$ tends to increase with width for quiescent prominences. In addition, the catalogue shows that high values of $t_d/P$ are reported, which means that our results are not as different as the observed results, just as we had thought. Moreover, strong damping of about $t_d/P\sim 1.25$ as the values obtained for longitudinal oscillations is the rule. As we found for horizontal transverse and longitudinal oscillations, \citet{2018ApJS..236...35L} did not find a correlation between $t_d/P$ and the filament dimensions.

Another interesting result that we can compare with observations is the variation of $P$ with height. \citet{2011A&A...531A..53H} studied the dependence of the periodicity and the attenuation on height of two successive trains of transverse oscillations observed on 2005 July 30. They found that the periods range between 90 and 110 min and increase slightly with height. Conversely, \citet{2014ApJ...795..130S} did not find differences of the period with the altitude in a transverse oscillation observed on 9 August 2012, so that they suggested that the prominence oscillated as a rigid body. These behaviours differ from our results and from those of \citet{2018ApJ...856..179Z}. We obtained that the period increases with height for longitudinal oscillations and decreases for transverse ones. These discrepancies suggest that a deeper analysis of the variations of $P$ with height is necessary. For example, \citet{2011A&A...531A..53H} reported long periods, and if we take our simulations into account together with the long periods observed in their study, we can suggest that the oscillations are longitudinal instead of transverse. In any case, knowing the oscillation polarisation with respect to the magnetic field is essential for understanding the nature of the observations. 

We have exhaustively analysed vertical oscillations for different scenarios. The origin of these oscillations is the relaxation process of our prominence model. For the stationary state, we induced a $v_z$ perturbation in the reference simulation to trigger vertical motions. The results (not shown in this work) reveal basically the same behaviour as the relaxation process. According to this, we could consider the vertical oscillations of the relaxation process as a winking filament if the velocity amplitude of the movement were sufficiently large. We did not take the nature of the disturbance into account here. For this reason, in a future work we will study the dependence of the oscillations varying the characteristics of the perturbation. We are interested in simulating oscillations triggered by a distant perturbation or by pressure pulses and changing the energy of the wave.

\begin{acknowledgements}
We acknowledge the support from grant AYA2017-85465-P (MINECO/AEI/FEDER, UE) and the Conselleria d'Innovació, Recerca i Turisme del Govern Balear to IAC\textsuperscript{3}. We are grateful to the ISSI Team led by Manuel Luna 'Large-Amplitude Oscillations as a Probe of Quiescent and Erupting Solar Prominences' for inviting us to be part of the Team. A.A. acknowledges the Spanish 'Ministerio de Economía, Industria y Competitividad' for the 'Ayuda para contratos predoctorales' grant BES-2015-075040. 
\end{acknowledgements}  

\bibliographystyle{aa}
\balance
\bibliography{references} 
\end{document}